\documentclass[aps,prmaterials,twocolumn,10pt,superscriptaddress]{revtex4-2}

\usepackage{amssymb,amsfonts,amsmath}
\usepackage{times}
\usepackage{graphicx}
\usepackage{color}
\usepackage{float}
\usepackage{cancel}
\usepackage{orcidlink}
\usepackage{hyperref} \hypersetup{colorlinks=true,citecolor=blue,linkcolor=blue,urlcolor=blue} % HYPERLINKS

\newlength{\figurewidth}
\newlength{\pagewidth}
\def\AFLOW{{\small AFLOW}} % acronyms go \small
\def\DFT{{\small DFT}}
\def\MLIAP{{\small ML-IAP}} \def\MLIAPs{{\small ML-IAP}s}
\def\MTP{{\small MTP}} \def\MTPs{{\small MTP}s}
 \def\PRAPs{{\small PRAP}s}
\def\MLIP{{\small MLIP}} \def\MLIPs{{\small MLIP}s}
\def\CSP{{\small CSP}} 
\def\GAP{{\small GAP}} \def\EDD{{\small EDD}}
\def\PES{{\small PES}} \def\PESs{{\small PES}s}
\def\EFS{{\small EFS}} \def\ALS{{\small ALS}}
\def\RMS{{\small RMS}} 
\def\RMSE{{\small RMSE}} \def\RMSEs{{\small RMSE}s}
\def\MAE{{\small MAE}} \def\MAEs{{\small MAE}s}
\def\RP{{\small RP}} \def\RPs{{\small RP}s}
\def\AP{{\small AP}}
\def\APRR{{\small AP-RR}}
\def\ARRR{{\small AR-RR}}

\begin{document}

\makeatletter \renewcommand\frontmatter@abstractwidth{\dimexpr\textwidth\relax} \makeatother % WIDEABSTRACT

\def\BUFFALO{\footnotesize Department of Chemistry, State University of New York at Buffalo, Buffalo, NY 14260, USA}
\def\MEMS{\footnotesize Department of Mechanical Engineering and Materials Science, Duke University, Durham, NC 27708, USA}
\def\CAMD{\footnotesize Center for Autonomous Materials Design, Duke University, Durham, NC 27708, USA}
\def\PSU{\footnotesize Department of Materials Science and Engineering, The Pennsylvania State University, University Park, PA 16802, USA}
\def\NCSU{\footnotesize Department of Materials Science and Engineering, North Carolina State University, Raleigh, NC 27695, USA}
\def\MST{\footnotesize Department of Materials Science and Engineering, Missouri University of Science and Technology, Rolla, MO 65409, USA}

\title{Machine Learned Interatomic Potentials for Ternary Carbides trained on the AFLOW Database}

\author{Josiah~Roberts\,\orcidlink{0000-0002-1359-5184}}\affiliation{\BUFFALO}
\author{Biswas~Rijal\,\orcidlink{0000-0003-3679-4191}}\affiliation{\BUFFALO}
\author{Simon~Divilov\,\orcidlink{0000-0002-4185-6150}}\affiliation{\MEMS}\affiliation{\CAMD}
\author{Jon-Paul~Maria\,\orcidlink{0000-0003-3604-4761}}\affiliation{\PSU}
\author{William~G.~Fahrenholtz\,\orcidlink{0000-0002-8497-0092}}\affiliation{\MST}
\author{Douglas~E.~Wolfe\,\orcidlink{0000-0002-3997-406X}}\affiliation{\PSU}
\author{Donald~W.~Brenner\,\orcidlink{0009-0009-1618-4469}}\affiliation{\NCSU}
\author{Stefano~Curtarolo\,\orcidlink{0000-0003-0570-8238}}\affiliation{\MEMS}\affiliation{\CAMD}
\author{Eva~Zurek,\orcidlink{0000-0003-0738-867X}} \email{ezurek@buffalo.edu}\affiliation{\BUFFALO}

\begin{abstract}
\noindent
Large density functional theory (\DFT) databases are a treasure trove of energies, forces and stresses that can be used to train machine learned interatomic potentials for atomistic modeling. Herein, we employ structural relaxations from the \AFLOW\ database to train moment tensor potentials (\MTPs) for four carbide systems: HfTaC, HfZrC, MoWC and TaTiC. The resulting \MTPs\ are used to relax $\sim$6300 random symmetric structures, and are subsequently improved via active learning to generate robust potentials (\RP) that can relax a wide variety of structures, and accurate potentials (AP) designed for the relaxation of low-energy systems. This protocol is shown to yield convex hulls that are indistinguishable from those predicted by \AFLOW\ for the HfTaC, HfZrC and TaTiC systems, and in the case of the CMoW system to predict thermodynamically stable structures that are not found within \AFLOW, highlighting the potential of the employed protocol within crystal structure prediction. Relaxation of over three hundred Mo$_{1-x}$W$_x$C stoichiometry crystals first with the \RP\ then with the \AP\ yields formation enthalpies that are in excellent agreement with those obtained via \DFT. 
\end{abstract}

\maketitle

\section*{\label{sec:Intro}Introduction}

Machine learned interatomic potentials (\MLIAPs), which are trained on density functional theory (\DFT) data, have irrevocably changed the way in which computational materials science is carried out. They have increased the complexity of the systems that can be studied via static calculations~\cite{LiuPei2023}, the duration of molecular dynamics trajectories~\cite{MLP_LongMD}, and made it possible to routinely model effects, such as anharmonicity~\cite{MTP+TILD}, which are commonly neglected due to the large computational cost associated with the requisite \DFT\ calculations. Moreover, they are becoming increasingly important in the field of crystal structure prediction (\CSP) where it may be necessary to relax hundreds or thousands of structures to find the global minimum and a handful of low-lying local minima~\cite{Tong2020}. As \CSP\ methods tackle systems with increased combinatorial complexity such as ternaries and quaternaries~\cite{Zurek:2020i}, it is becoming increasingly important to develop protocols that can be used to train and employ \MLIAPs\ for \CSP.  

Some of the most well-known \MLIAPs\ include neural networks (NNs) ~\cite{Behler-Parrinello,Hajinazar2017},  the spectral neighbor analysis potential ({SNAP)}~\cite{Thompson:2015a}, moment tensor potentials (\MTPs)~\cite{Shapeev:2016a},  the Gaussian approximation potential (\GAP)~\cite{Bartok:2010a}, and, more recently ultra-fast (UF)~\cite{xie2023ultra}, and ephemeral data-derived (\EDD) potentials~\cite{Pickard2022}. Provided an \MLIAP\ can accurately describe the potential energy surface (\PES) of a multi-component system, it can significantly accelerate a \CSP\ search while at the same time enabling exploration of wider regions of compositional space~\cite{Tong2020}. Indeed, recent algorithmic developments have shown promising results: \CSP\ methods including random and evolutionary searches, as well as particle swarm optimization have been coupled with various \MLIAPs\ and applied to predict crystalline structures of boron~\cite{Yang:2021a,Podryabinkin:2019a,Deringer:2018a}, carbon~\cite{Deringer:2017a,Podryabinkin:2019a}, sodium~\cite{Podryabinkin:2019a}, phosphorus~\cite{Deringer2020}, lithium~\cite{Ma-Li:2023} as well as Mg-Ca~\cite{MgCa-NN} and various metal-tin~\cite{LiSn-NN,Thorn} alloys at 1~atm and under pressure. For ternary crystalline systems, early work showed that adaptive classical potentials can accelerate \CSP\ in the Mg-Si-O system~\cite{adaptive:2014a}, and recently ephemeral deep learning potentials combined with random searching have been used to predict candidate structures for a reported Lu-N-H superconducting phase~\cite{Ferreira}, and Zn(CN)$_2$ metal-organic-frameworks~\cite{Salzbrenner}. 

A challenge in developing \MLIAPs\ for \CSP\ is that they need to be able to predict the energies of unstable structures, as well as those that are close to the local minima in the \PES. Moreover, because \MLIAPs\ are poor at extrapolation, they cannot accurately predict the energies (or forces or stresses) of configurations that differ significantly from those they have seen before. Traditionally, \DFT\ data sets containing thousands to hundreds-of-thousands of structures have been generated for the training and testing of \MLIAPs~\cite{Podryabinkin:2019a,Gubaev:2019a}. A wide range of procedures have been employed to create these \DFT\ data sets including generating structures randomly~\cite{Podryabinkin:2019a}, perturbing the geometries of such structures via `shaking'~\cite{Pickard2022}, \emph{ab initio} molecular dynamics runs at various temperatures~\cite{Zuo}, decorating predefined lattices with atoms of different types while simultaneously varying the chemical composition~\cite{Gubaev:2019a}, relaxation of structures generated via constrained evolutionary searches~\cite{Hajinazar2017}, straining crystal lattices, creating defect structures, and more~\cite{Zuo,liu:2024}.

Unfortunately, even when large \DFT\ data sets are employed for training, it is unlikely that the resulting \MLIAPs\ can predict, with sufficient accuracy, the energies of the various structures encountered in the course of a \CSP\ run. One strategy that has been proposed for the generation of a multi-purpose \MLIAP, given a  limited number of \DFT\ calculations, relies on the automatic iterative building of the fitting database by selecting the most diverse structures ~\cite{Bernstein:2019b}. Additional techniques include various active learning~\cite{Smith:2018a,Zhang:2019a} and learning-on-the-fly~\cite{Jinnouchi:2019a} methods, where the potential is updated and improved during the course of the search. It has been suggested that active learning could be used to generate two \MLIAPs: a robust one that is able to optimize any structure the \CSP\ algorithm encounters and make rough predictions, and an accurate one trained on (and able to relax) only the low-energy structures~\cite{Podryabinkin:2019a, Gubaev:2019a}. 

\MLIAP\ based simulations where the potential is updated on-the-fly may be initialised using either an empty/untrained potential~\cite{Gubaev:2019a}, or one that has been pre-trained. The former strategy may require just as many, if not more, \DFT\ evaluations than the latter because the likelihood of encountering a configuration that is deemed extrapolative is high, necessitating a retraining of the \MLIAP~\cite{Podryabinkin:2019a}. Thus, there appears to be `no-free-lunch' since the construction of a reliable \MLIAP\ requires numerous expensive \DFT\ calculations. At the same time, a large number of databases exist -- \AFLOW\ (Automatic {\small FLOW})~\cite{AFLOW,curtarolo:art190,curtarolo:art191}, the Materials Project (MP)~\cite{matproj}, the Open Quantum Materials Database~\cite{Saal:2013}, etc.\ -- each containing millions of \DFT\ evaluations of the energies, forces and stresses of extended systems. Furthermore, it is becoming standard practice for researchers to deposit the \DFT\ data generated during the course of a computational project in repositories such as {\small NOMAD}~\cite{Draxl:2019}, {\small OCELOT}~\cite{Risko:2021} and {\small NIST} Materials Data~\cite{NISTmats}. One way this data is being used is to train \MLIAPs\ for the computational study and exploration of the vast PES of all possible chemistries. Some of the forefront examples of such `universal' \MLIAPs , which can predict energies, forces and stresses using equivariant graph neural networks, include M3GNet~\cite{m3gnet}, CHGNet~\cite{chgnet}, ALIGNN-FF~\cite{alignn}, MACE-MP-0~\cite{MACE}, and GNoME~\cite{GoogUniverse}.  

The training of potentials on already existing \DFT-data is illustrated here by combining outputs present within the \AFLOW~\cite{AFLOW} database with \MTPs ~\cite{Shapeev:2016a}, as implemented within the Machine Learned Interatomic Potentials (\MLIP) program~\cite{MLIP2}. Specifically, chemically sensible structures, which are randomly selected from the relaxation trajectories stored within the \AFLOW\ database, are used to train an \MTP\ that is subsequently employed to relax a large number of random symmetric structures spanning a wide composition range. In contrast to the recently developed universal ML-IAPs~\cite{m3gnet,chgnet,alignn,MACE,GoogUniverse,GoogHarmPot},  here we only train on a subset of the data found within \AFLOW , chosen for the application in-mind. Therefore, the \MLIAPs\ developed here are system-specific, and not universal. What distinguishes our study from prior works~\cite{Podryabinkin:2019a,Pickard2022,Zuo,liu:2024,Hajinazar2017} is that rather than generating our own DFT-training set, we employ already existing data found within \AFLOW. In a further step, the \AFLOW-trained potentials are improved via active learning, generating \MLIAPs\ that can be robust (able to roughly optimize any configuration) or accurate (for the optimization of only low-energy structures near the convex hull). Thus, only a small number of supplementary \DFT\ calculations are required to develop system-specific \MLIPs\ that enable the computational exploration of evermore complex \PESs\ towards the discovery of materials. 

A utility package that automates this training process, the \emph{Plan for Robust and Accurate Potentials} (\PRAPs), is described.  The method is used to determine the zero-Kelvin phase diagrams of four ternary metal carbides, chosen because they represent materials with superlative mechanical properties~\cite{Maria1,Maria2}. The \PRAPs\ generated convex hulls whose lowest energy structures are optimised with \DFT\ are in good agreement with the DFT-relaxed \AFLOW\ hulls. Moreover, in the case of the CMoW system, thermodynamically stable structures are found that are not present in the \AFLOW\ data. Further calculations with the accurate potential find a variety of Mo$_{1-x}$W$_x$C stoichiometry phases at/near the tie-line indicating the possibility of a solid-solution with a very low miscibility gap critical temperature.

\section*{Results and Discussion}\label{sec:rd}
\subsection{Plan for Robust and Accurate Potentials (PRAPs)} \label{sec:work}

%\noindent{\bf Plan for Robust and Accurate Potentials (\PRAPs).} 
We created \MTPs\ of varying complexity for a number of ternary metal carbides and investigated their capabilities to predict structures outside of their training sets. Our choice of \MTPs\ was motivated by their excellent balance between model accuracy and computational efficiency~\cite{Zuo,HartNPJ}, their application towards multicomponent systems~\cite{MTP+HEA,Gubaev:2019a, gubaev2023performance,liu:2024,zeng:2024}, their ability to predict phonons and thermodynamic properties,~\cite{TherMTP,MTPhonons,MTP+TILD} and the availability of a powerful active learning scheme (\ALS) that has been interfaced with the \MTP\ method~\cite{MLIP1,Gubaev:2019a}. 
\begin{figure*}
    \centering
    \includegraphics[width=0.95\textwidth]{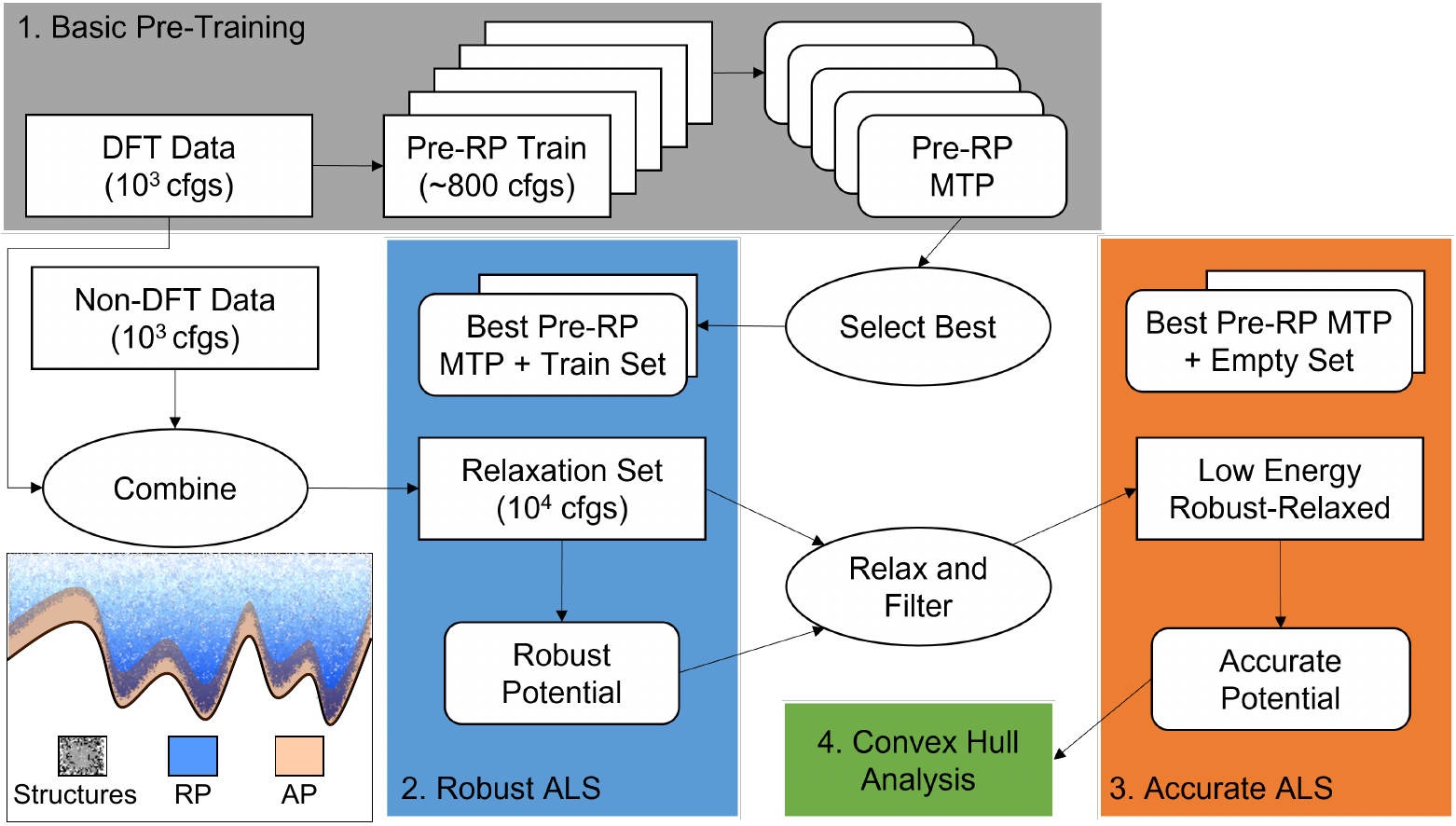}
    \caption{\small Workflow in the \emph{Plan for Robust and Accurate Potentials} (\PRAPs) package, which automates the generation of a moment tensor potential (\MTP), given any quantum mechanical training set (here the online \AFLOW\ database~\cite{AFLOW}). Five \MTPs\ are trained on a set of configurations (cfgs) and the best (the pre-Robust Potential, pre-\RP) is chosen (gray box). The pre-\RP\ is employed to initialise the training of a \RP\ via an active learning scheme (\ALS, blue box). The \RP\ is generated by relaxing the configurations comprising the Relaxation Set. The \RP-relaxed lowest energy configurations are chosen (filtration step) to train an Accurate Potential (\AP) via active learning (orange box). The \AP-relaxed structures may be sent for relaxation via \AFLOW-\DFT, followed by subsequent analysis (green box). The inset schematically illustrates the energy distribution of the structures that can be relaxed with the \RP\ and the \AP. Optionally, structures generated using such a procedure may not be exact since the intermediate structures in a DFT relaxation trajectory may have erroneous energies or forces resulting from Pulay stress that arises  when the lattice vectors are varied.   another procedure (here \textsc{RandSpg}~\cite{RandSPG}), labelled as `Non-\DFT\ Data', may be relaxed via the \RP\ and/or the \AP, and this auxillary set of structures also comprises the Relaxation Set. Rectangles with sharp (curved) corners represent structural data (potentials). }
    \label{fig:workflow}
\end{figure*}

The \MLIP\ software package trains \MTPs\ and uses them to relax the geometries and minimize the energies of a wide variety of chemical systems~\cite{MLIP2}. The simplest form of training, basic training, employs the energy, force and stress (\EFS) data of a set of configurations, as obtained from \DFT\ or other quantum chemical calculations, to generate an \MTP. The complexity of the \MTP\ is described by a user-selected \emph{level}, a notation containing information about the number of basis functions and parameters comprising the potential. The \ALS\ employs a D-optimality criterion to calculate the extrapolation degree or grade, $\gamma$, for every structure that is generated throughout the course of the simulation (relaxation trajectory or molecular dynamics run)~\cite{MLIP1,Gubaev:2019a}. \MLIP\ automatically selects configurations to be added to the training set if their $\gamma$ falls within a user-defined range; we choose the default $2 < \gamma \le 10$. The calculation (relaxation or molecular dynamics run) is terminated if $\gamma$ exceeds the upper bound, triggering retraining of the \MTP, and the procedure repeats until the simulation finishes with $\gamma \le 10$. Further information about \MTPs,\ including their functional form, the quantities included in the cost function, and details of the active learning procedure are provided in Section S1 of the Supplementary Information.

In what follows we give a brief overview of the \PRAPs\ utility package employed in this study; a forthcoming manuscript will describe its composition and usage. \PRAPs\ interfaces with \MLIP\ and automates the creation of a Robust Potential (\RP) and Accurate Potential (\AP) using the aforementioned \ALS, and employs basic training for supplemental tasks. From a given a set of configurations obtained from AFLOW, a subset of $\sim$800 configurations is chosen randomly and used to train an \MTP\ (Figure \ref{fig:workflow}, gray box). This procedure is repeated five times, generating five different \MTPs\ (as recommended in the \MLIP\ manual~\cite{MLIP2}), mimicking a cross-validation procedure. From these \PRAPs\ finds the `best' \MTP, defined as the one that identifies the most high-and-low-energy structures (ten of each are compared), or if multiple \MTPs\ fulfill this criteria, the \MTP\ with the lowest root-mean-squared (\RMS) training error. This step is intended to mitigate the initial random parameters that \MLIP\ assigns at the beginning of training, but users may bypass it if they so desire. To ensure that training is performed on sensible structures, \PRAPs\ can filter out configurations with undesirable interatomic distances (default), or cell volumes. If relaxation trajectories comprise the data set, dozens of ionic steps originating from the same system may be present. Though using many similar configurations for training may be beneficial in some cases, in others it may be desirable to exclude the intermediate steps and only include the final relaxed configuration, and \PRAPs\ provides an option for users to select the desired behaviour. 

The \RP\ training (Figure \ref{fig:workflow}, blue box) begins using the \PRAPs-determined `best' \MTP.  The set of configurations may be optionally augmented with structures lacking any \EFS\ data (Figure \ref{fig:workflow}, left-hand-side, no box), which are combined with the initial \DFT\ data set to form the relaxation set. The relaxation set is used to train the \RP\ by active learning. The active learning begins with the `best' training set and \MTP\ from the pre-training step (if performed, otherwise \PRAPs\ trains from scratch). During the active learning process, the relaxation set is relaxed using \MLIP's built-in functionality and new structures are added to the training set based on the D-optimality criterion (a detailed discussion of this step is available in Section S1). The final output is a potential capable of performing a reasonable relaxation of most any configuration in the relaxation set.

After relaxation with the \RP, \PRAPs\ filters out structures with energies above a specified cutoff (default 50 meV/atom) for each composition present. This creates the `Robust Relaxed' set, which is employed to train the \AP\ via active learning (Figure \ref{fig:workflow}, orange box). The procedure for training the \AP\ is similar to the \RP\ except that \PRAPs\ begins with an empty training set instead of the `best' pre-trained \MTP's training set. We find that this results in better predictions as the \AP\ should not see non-ground-state structures during training. The resulting \AP\  should improve the \EFS\ predictions for these low-energy structures. For additional efficiency, users can alter the convergence criteria of the \AP\ training at the cost of a few meV/atom in the training error (as described more thoroughly below).

\PRAPs\ performs a number of analyses throughout and at the end of the training procedure. In addition to calculating the training and prediction \RMSE\ and mean absolute error (\MAE), and comparing the ten high-and-low-energy structures, it can generate a set of convex hull candidates, invoke \AFLOW~\cite{AFLOW, AFLOWstd} to relax them, and use this data to generate composition-energy convex hull plots (Figure \ref{fig:workflow}, green box). Finally, \PRAPs\ contains a checkpoint system to allow users to start or re-start a job from a certain step. \PRAPs\ is primarily a project management software that performs and automates many menial tasks, and generates plots that may be desired, thereby reducing the amount of human time required to generate \MTPs\ for a particular system and analyze their performance.

In what follows we seek to answer two questions using \PRAPs: (i) Can already existing quantum-mechanical data be used to train an \MTP\ that is able to reproduce the DFT-calculated convex hull for a particular ternary alloy? (ii) And, can we use these \MTPs\ to discover thermodynamically stable structures comprising the convex hull that are not present in the \DFT-training set? In what follows, we show the answers to both questions is `yes'. \\

\subsection{Machine Learned Interatomic Potentials from \AFLOW\ Data}

%\noindent{\bf Machine Learned Interatomic Potentials from \AFLOW\ Data.}
To illustrate how already-existing \DFT\ results can be scraped from large databases and repurposed towards the generation of system specific \MLIPs, we chose four ternary alloy systems: CHfTa, CHfZr, CTaTi, and CMoW. TaC, HfC, ZrC, and TiC all adopt the rocksalt structure and are refractory ceramic materials with desirable mechanical properties~\cite{TiC_Nakamura}. Though MoC can adopt the same rocksalt structure under pressure~\cite{CMo_Clougherty}, at ambient conditions MoC and WC prefer a hexagonal arrangement instead~\cite{CMoW_Schuster}. For nearly a century the propensity for transition metal carbides to form high-melting point solid solutions with compositions such as Hf$_{1-x}$Ta$_x$C has been known~\cite{Harrington2019,Vorotilo2022}. The metal formulation can be engineered to contain multiple atoms, and when there are five or more metals, the configurational entropy stabilizes single-phase high entropy carbides (HECs)~\cite{Maria1,Sarker:2018a,DEEDetal} and their thin films~\cite{Maria2}.  Recently, \MLIAPs\ have been developed for various HECs including a deep learning potential for  (ZrHfTiNbTa)C$_5$~\cite{dai2020theoretical} and a low rank potential for  (TiZrNbHfTa)C$_5$~\cite{pak2023machine}. 
Herein, we illustrate that data found within \AFLOW, coupled with active learning, can be used to train \MTPs\ for ternary carbides, with the future goal of HECs in mind. 

For training we employed \DFT\ relaxation trajectories of $\sim$210 structures obtained from the \AFLOW\ database, which have been generated through a combination of structure prototyping of naturally occurring compounds~\cite{Ref14, Ref15, Ref16} and structure enumeration algorithms~\cite{Ref13}, followed by relaxation using density functional theory. This resulted in $\sim$5500-6000 individual configurations on which the \MTP\ was trained (Table \ref{table:cfg}). From this training set, $\sim$800 configurations, with a minimum interatomic distance greater than 1.1~\AA{}, were chosen randomly and \MTPs\ of levels 10, 16 and 22 were trained in the ``basic'' mode. For the CMoW system, no level 22 data is reported due to the excessive computational cost required to obtain well-trained potentials. Five trainings were performed, and the best potential was chosen as the pre-Robust Potential (pre-\RP). As expected, the training errors for the pre-\RP\ (Table S1) decreased with increasing \MTP\ level, with the average \MAE\ (\RMSE) on the energies being 27 (44), 16 (25) and 8 (13)~meV/atom, and for the forces being 75 (217), 51 (147) and 26 (78)~meV/ \AA{} for levels 10, 16 and 22, respectively. Comparison of the ten highest and ten lowest energy structures as predicted by \DFT\ and the pre-\RP\ revealed that the \MTP\ rarely miscategorized the structures, but had difficulty correctly categorizing more than 7 in the correct highest or lowest energy set (Table S2).

\begin{table}[t]
    \centering
    \begin{tabular}{|c|c|c|c|c|}
        \hline
        System & \AFLOW\ structures & Configurations & \textsc{RandSpg} structures \\ 
        \hline \hline
        CHfTa & 209 & 6064  & 6346 \\
        \hline
        CHfZr & 210 &5811  & 6356 \\
        \hline
        CMoW & 210 & 5524  &6338 \\
        \hline
        CTaTi & 211 & 5831  & 6347 \\
        \hline
    \end{tabular}
    \caption{\small Number of unique structures for the studied ternary carbides found within the \AFLOW\ database, and the number of individual ionic steps (configurations) from their structural relaxations. The number of individual structures generated by \textsc{RandSpg} that were employed to develop the robust and accurate potentials using the \MLIP\ program package are also provided.}
    \label{table:cfg}
\end{table}

Previous studies have suggested that when used for \CSP, \MTPs\ should be trained on a very diverse set of structures~\cite{Podryabinkin:2019a, Gubaev:2019a}, and as Table~\ref{table:cfg} shows, the original \AFLOW\ data was somewhat limited. To obtain this diversity, we used the \textsc{RandSpg}~\cite{RandSPG} program to create crystal lattices with up to eight atoms in the unit cell, for all possible ternary compositions. Lattice vectors were constrained to fall between 3-10~\AA{}, and unit cell volumes between 200-600~\AA{}$^3$, with a minimum interatomic distance of 1.1~\AA{}. For each ternary system, $\sim$6300 structures were generated (Table \ref{table:cfg}) and combined with the initial \AFLOW\ data to form the relaxation set (Figure \ref{fig:workflow}) used in training the \RP. 

For a given stoichiometry \textsc{RandSpg} determines the compatible spacegroups, based upon the Wyckoff positions, and randomly chooses one prior to decorating its sites with atoms, thereby enabling the creation of random symmetric crystal lattices. Employing symmetric structures in the first generation of an evolutionary or particle-swarm directed \CSP\ search greatly decreases the number of configurations that need to be optimized to locate the global minimum in the \PES. This can be understood by noting that symmetric structures tend to be either very stable or unstable, spanning a greater amount of the potential energy hypersurface than those generated without symmetry constraints~\cite{Zurek:2020i}. Indeed, tests have shown that the average energy of random structures that are symmetric is higher (less negative) than those that are purely random~\cite{RandSPG}.

\begin{table*}[t]
\centering
\begin{tabular}{ |c|c|c|c|c| }
 \hline
 Alloy & \multicolumn{2}{|c|}{Energy Errors (meV/atom)} & \multicolumn{2}{|c|}{Force Errors (meV/\AA{})} \\
 \hline \hline
 Level 10 & Robust Potential & Accurate Potential & Robust Potential & Accurate Potential \\
 \hline
 CHfTa & 45 (75) & 24 (36)  & 89 (281) & 82 (270) \\
 \hline
 CHfZr & 53 (87) & 37 (62) & 81 (255) & 57 (170) \\
 \hline
 CMoW & 71 (125) & 71 (93) & 120 (377) & 50 (160) \\
 \hline
 CTaTi & 51 (85) & 53 (70) & 99 (303) & 76 (220) \\
 \hline \hline
 Level 16 & Robust Potential & Accurate Potential & Robust Potential & Accurate Potential \\
 \hline
 CHfTa & 42 (72) & 29 (90) & 79 (238) & 53 (169) \\
 \hline
 CHfZr & 43 (77) & 22 (30) & 68 (211) & 42 (112) \\
 \hline
 CMoW & 68 (118) & 168 (214) & 113 (359) & 96 (276) \\
 \hline
 CTaTi & 41 (60) & 45 (59) & 90 (285) & 56 (161) \\
 \hline \hline
 Level 22 & Robust Potential & Accurate Potential & Robust Potential & Accurate Potential \\
 \hline
 CHfTa & 44 (70) & 21 (57) & 85 (270) & 23 (71) \\
 \hline
 CHfZr & 39 (62) & 29 (91) & 61 (197) & 18 (59) \\
 \hline
 CTaTi & 36 (58) & 82 (218) & 73 (223) & 25 (79) \\
 \hline \hline
Level 16$^*$ & Robust Potential & Accurate Potential & Robust Potential & Accurate Potential \\
 \hline
 CHfTa & 156 (126) & 16 (19) & 129 (381) & 50 (150) \\
 \hline
 CHfZr & 107 (78) & 24 (37) & 93 (278) & 40 (114) \\
 \hline
 CMoW & 155 (97) & 67 (120) & 108 (336) & 52 (196) \\
 \hline
 CTaTi & 153 (119) & 92 (112) & 110 (345) & 45 (134) \\
 \hline 
 \end{tabular} \\
 $^*$Calculations performed without pre-training the \MTP\ on the \AFLOW\ data (grey box in Figure \ref{fig:workflow}).
\caption{\small Prediction errors for energies (meV/atom) and forces (meV/$\AA{}$) by system, level, and potential. Each box contains the mean-absolute-error and, in parentheses, the root-mean-squared errors (\RMSE). Only the Robust Potential (\RP) and the Accurate Potential (\AP) developed during the active learning scheme (\ALS) as illustrated in Figure \ref{fig:workflow} are shown here; data for the pre-Robust Potential can be found in the SI. The training errors for each potential reflect the \RMS\ values obtained by comparing the energies of that potential's training set as obtained by single-point \DFT\ calculations against those predicted by the respective potential. The prediction errors for the \RP\ indicate comparison of the energies of the original \AFLOW\ data with the energies predicted by the respective potential. The prediction errors for the \AP\ indicate comparison of only the low-energy \AFLOW\ structures with those predicted by the respective potential. }
\label{table:rmse}
\end{table*}

Table \ref{table:rmse} provides the prediction errors for the energies and forces for both the \RP\ and the \AP\ for different \MTP\ levels. Determining the prediction error when such active learning schemes are used is complicated by the fact that each level of theory (\DFT\ or \MTP\ level) will encounter different structures during the relaxation process. While it is possible to predict the energy of every structure comprising the \AFLOW\ relaxation trajectories via the \RP\ or the \AP, such a procedure is fraught with difficulties because of the noise in the \DFT\ relaxation trajectory data. Specifically, the relaxation trajectory may have erroneous energies, forces and stresses resulting from a change in the planewave basis as the lattice vectors are varied during the course of the optimization. Moreover, the \AFLOW\ data contained a few configurations whose magnitude of per-atom-energies were substantially larger than others, though they survived the distance-filtering-criteria (Supplementary Figure 1. Therefore, in Table \ref{table:rmse} we have opted to compare the \DFT\ energies and forces on the final \AFLOW\ relaxed structures, which are accurately calculated, against their \RP\ and \AP\ predicted values.

A number of trends can be observed from examination of Table \ref{table:rmse}. First, the \MAE\ are notably smaller than the \RMSE, suggesting the presence of poorly-predicted outliers. The prediction errors with the \RP\ generally decrease with increasing level of \MTP, but the \MAE\ do not fall below 40~meV/atom for energies and 60~meV/\AA{} for the forces. While these errors might seem large compared to the $<4$~meV/atom and $<160$~meV/\AA{} \RMSEs\ computed for single-component systems wherein the testing and training dataset contained structures that could be derived via perturbations of the ground state crystal~\cite{Zuo}, they are in-line with some of the errors presented in Ref.\ \cite{liu:2024} that trained \MTPs\ for Li-Al alloys and applied them to a broad range of compositions and crystal lattices. In Ref.\ \cite{Gubaev:2019a} the only errors reported for the Al-Ni-Ti system were for the training set with \MAEs\ (\RMSEs) of 18 (27)~meV/atom for the \RP\ and 7 (9)~meV/atom for the \AP . In Ref.\ \cite{Podryabinkin:2019a} a \RP\ yielded a training \RMSE\ of 170~meV/atom for allotropes of boron, and an \AP\ yielded errors of 11~meV/atom for the 100 lowest-energy structures that were found.

A key difference between the workflow presented here as compared to previous work by Gubaev and co-workers~\cite{Gubaev:2019a} is the pre-training step on the \AFLOW\ data. To test what effect this may have, we performed the \PRAPs\ procedure starting from an empty \MTP, that is one that had not been trained on \AFLOW\ data for a level 16 \MTP. Therefore, the basic pre-training denoted schematically in the grey box in Figure \ref{fig:workflow} was not performed. Table \ref{table:rmse} clearly illustrates that the \MAEs\ for the energies were significantly improved by the \AFLOW\ pre-training, and the \RMSEs\ were slightly improved (by 94 and 16~meV/atom, on average).

A common practice in \CSP\ is to use less accurate, but quicker methods to estimate the energies of a large number of structures, and then to optimize those with low energies with progressively more accurate, but costlier, methods. This workflow saves computational expense by filtering out structures that are unlikely to be stable prior to performing calculations that yield a more predictive rank order. A similar strategy was employed in Refs.~\cite{Gubaev:2019a} and~\cite{Podryabinkin:2019a} where both robust and accurate \MTPs\ were trained for \CSP. In addition to generating 375,000 binary and ternary bcc, fcc and hcp-type unit cells, 1463 Al-Ni-Ti ternary structures were created via decorating prototypes~\cite{Gubaev:2019a}. It was noted that the prototype-derived structures might possess geometries that were not chemically sensible because of short metal-metal distances and too-small volumes, which could lead to large \MTP\ prediction errors. To circumvent this problem, structures with formation energies (as determined with a Robust \MTP) that were within 100~meV/atom of the convex hull were chosen for re-relaxation using active learning starting from an empty \MTP\ to generate an \AP. In another related work, the large training error of 170~meV/atom obtained in a \CSP\ search performed on boron, which has a very complex \PES, prompted the authors to retrain the \MTP\ on low-energy structures so it could accurately predict their energies~\cite{Podryabinkin:2019a}. 

Similar to these two studies~\cite{Gubaev:2019a,Podryabinkin:2019a}, our \textsc{RandSpg} generated structures span a wide energy range. Although the \RPs\ obtained from including them in the training set have large errors, they are able to correctly identify most low-energy configurations, and not misidentify them as having high energies. Therefore, by choosing the \RP-relaxed structures whose energies were within 50~meV/atom of the minimal energy structure for each composition, we can curate a new set of training data that can then be used to create an \AP.  Because the low-energy structures are expected to be better behaved than the full relaxation set, possessing nearly optimized geometries that are chemically sensible, the \AP\ should yield lower training and prediction errors than the \RP. 

In many cases both the training and prediction \RMS\ errors obtained for the \AP\ decreased with increasing \MTP\ level, however they did not fall below 50~meV/atom. Considering that most high-throughput and \CSP\ searches are performed in a stepwise procedure, at each level filtering undesirable structures prior to increasing the accuracy by which the remaining subset is treated, this error is acceptable. Indeed, in both Refs.\cite{Gubaev:2019a} and~\cite{Podryabinkin:2019a} all structures with sufficiently low energy, as predicted by the \AP, were re-optimized with \DFT\ to obtain an improved energy ordering. Similarly, in \CSP\ searches on Lu-N-H~\cite{Ferreira} and Sc$_x$H$_y$~\cite{Salzbrenner} the lowest energy structures, as predicted by an \MLIAP, were relaxed with \DFT\ to determine a more accurate energy rank.

For an \MTP\ level of 22, the training in some cases took substantially longer than at lower levels. The reason for this is that the \ALS, as originally designed~\cite{MLIP1}, terminates when \MLIP\ does not find any configurations that need to be added to the training set. Here a different protocol was used, where the active learning for the \AP\ was stopped when the number of structures to be added to the training set was less than 1\% of its size. This choice was motivated by the observation that at times fewer than ten configurations were being added to a training set of thousands in a single iteration, which took a full day to process, for a gain in \RMS\ training error that was $<$~1~meV/atom. Tests showed that the choice of a variety of early termination criteria (which can be chosen as options in \PRAPs) comes at the cost of 2-5~meV/atom in training error. 

\begin{figure}
    \centering
    \includegraphics[width=0.5\textwidth]{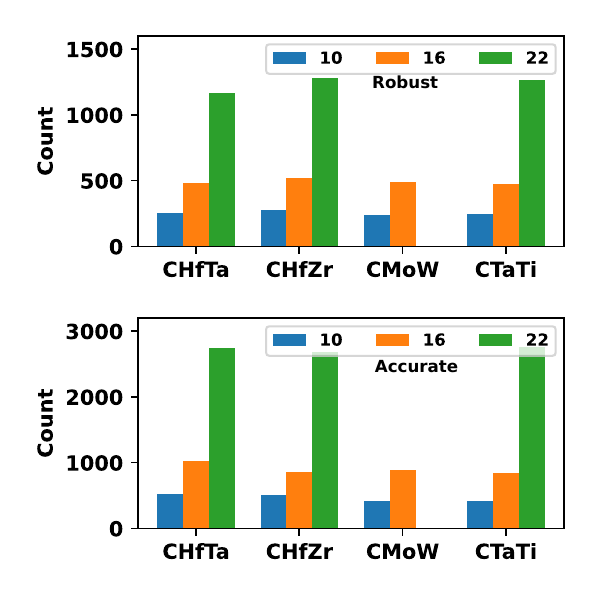}
    \caption{\small Number of density functional theory (\DFT) single point energy evaluations performed during the generation of the robust (top) and accurate (bottom) potentials according to the procedure illustrated in Figure \ref{fig:workflow} for each ternary carbide system at a particular \MTP\ level (see legend).}
    \label{fig:calcs}
\end{figure}

A key question to be answered is: ``Does the ML-aided procedure reduce the total computational expense?'' For each ternary the \AFLOW\ and \textsc{RandSpg} datasets contained $\sim$210 and  $\sim$6300 individual structures, respectively (Table \ref{table:cfg}). Therefore, $\sim$6500 geometry optimizations would be needed to relax all of these configurations. Using \MTPs, a maximum number  of single point \DFT\ energy evaluations performed during the course of the training was $\sim$3000 (Figure \ref{fig:calcs}) for the CTaTi system at level 22. Dividing the number of configurations comprising the \AFLOW\ data by the number of structures gives an average estimate of the number of steps required per geometry optimization ($\sim$28). Thus, our protocol reduces the number of \DFT\ evaluations that would need to be performed to relax all of the considered structures by a factor of $\sim$30 or more (as high as 180 for the CHfTa system at an \MTP\ level of 10).  
We note that as \MTP\ level increases, the number of single-point calculations increases (as does the total training time), as expected (\emph{e.g.}\ see Table 3 in Ref.\ \cite{MLIP2}). The reason for this is that higher level \MTPs\ contain more parameters, and therfore require more training data to fit these parameters.

Finally, we consider a few miscellaneous points about the setup and training procedures. Since the training data was predominantly for cubic structures, the HEAs which inspired our choice of elements typically crystallize in a cubic manner~\cite{Maria1, Maria2}, and the ternary carbides considered herein assume either cubic or hexagonal lattices, it is fair to ask if the results could be improved if the \textsc{RandSpg} structures were generated using only spacegroups with these symmetries. A test on all four systems, at an \MTP\ level of 10, revealed that both training and prediction errors were very similar when using a reduced set of $\sim$200 \textsc{RandSpg} structures across only HCP, FCC, and BCC-type unit cells in addition to the \AFLOW\ data (Table S3). A reader may ask why the pre-training step is carried out here since active learning can be performed on an empty training set~\cite{MLIP2, Gubaev:2019a, Podryabinkin:2019a}. Another test on CHfTa at level 10, showed that omitting the pre-training step resulted in a savings in time at the cost of training and prediction error (Table S4). The user can keep these trade-offs in mind when designing the computational protocol to be employed for their specific situation. 

%\subsection{Predicted Convex Hulls and Solid Solution Forming Ability}

\subsection{Predicted Convex Hulls} 
\begin{figure*}
    \centering
    \includegraphics[width=0.95\textwidth]{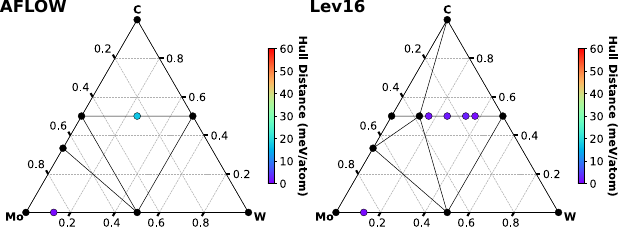}
    \caption{\small Convex hulls obtained by DFT-optimizing structures predicted to be within 50~meV/atom of the convex hull obtained from \AFLOW\ (left) and the \PRAPs\ procedure at an \MTP\ level of 16 (right) for the CMoW system. The black dots are structures lying on the convex hull, and phases within 60~meV/atom of the hull are represented by color dots (see color bar). Purple dots are within 1~meV/atom of the hull.}
    \label{fig:cmow}
\end{figure*}

Let us now examine the convex hulls and investigate the structures that \PRAPs\ relaxed with the robust and accurate potentials. Since optimization of random symmetric configurations is the first step in \CSP, we examined if the aforementioned workflow could discover lattices not found within \AFLOW\ whose energies lie on, or close to the convex hull. We filtered out the \AP\ predicted configurations that were within 50~meV/atom of the lowest energy structure for each composition and DFT-relaxed them via \AFLOW. The resulting geometries were then concatenated with the fully relaxed \AFLOW\ data to produce convex hulls, for different \MTP\ levels (Figure \ref{fig:cmow} and Figures S9 and S10).

\begin{figure*}
    \centering
    \includegraphics[width=0.95\textwidth]{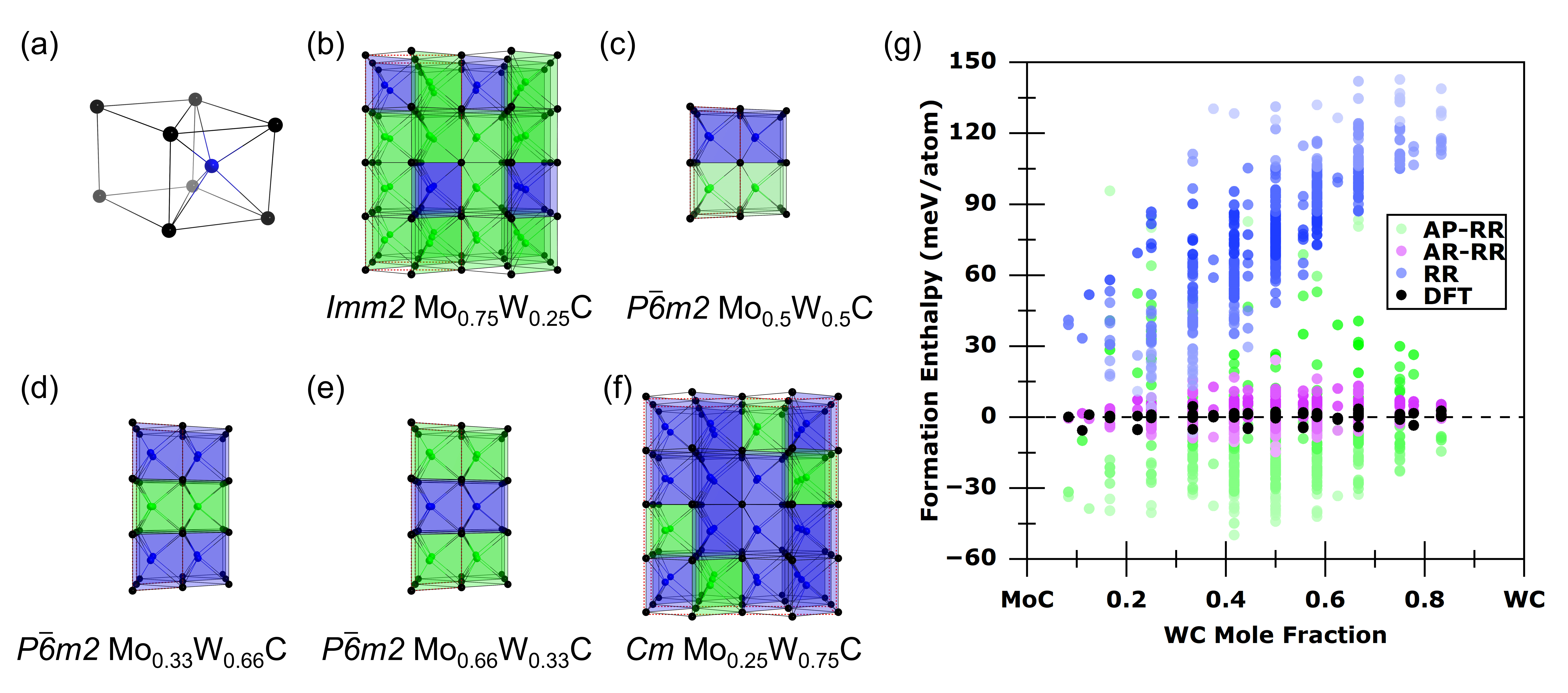}
    \caption{\small Crystal structures of (a) parent phases: $P\bar{6}m2$ WC (isostructural with MoC), (b) $Imm2$ Mo$_{0.75}$W$_{0.25}$C, (c) $P\bar{6}m2$ Mo$_{0.5}$W$_{0.5}$C, (d) $P\bar{6}m2$ Mo$_{0.33}$W$_{0.66}$C, (e) $P\bar{6}m2$ Mo$_{0.66}$W$_{0.33}$C and (f) $Cm$ Mo$_{0.25}$W$_{0.75}$C. Carbon atoms are colored black, tungsten atoms are blue and molybdenum atoms are green. Colored polyhedra are employed to emphasize the decoration of the structure by the two types of metal atoms. The $c$-axis is oriented perpendicular to the metal/carbon triangular nets. (g) The formation enthalpy, $\Delta H$, in meV/atom for the reaction $(1-x)\text{MoC}+ x(\text{WC}) \rightarrow \text{Mo}_{1-x}\text{W}_x\text{C}$ as a function of MoC/WC composition.  \DFT\ values (black dots) are provided along with results obtained following relaxation with the robust potential ({\small RR}), prediction of the energetics of the RR structures with the accurate potential (\AP-{\small RR}), and further relaxation of the RR structures with the accurate potential (\ARRR) given by blue, green and pink dots, respectively.}
    \label{fig:cmow_graph}
\end{figure*}

The data found within \AFLOW\ for the CHfTa, CHfZr, and CTaTi systems all contained multiple compounds on or near the convex hull with the M$_{1-x}$N$_x$C stoichiometry ($x=0.25, 0.33, 0.5, 0.67, 0.75$) (Figures S3-S13). Examination of these structures suggested that they could all be obtained via relaxation of metal-carbide rocksalt structures whose metal sites were decorated with two different types of atoms, as would be expected for a solid-solution. On the other hand, for the CMoW system, \AFLOW\ did not contain any ternary carbide phases with a Mo$_{1-x}$W$_x$C composition that were on the hull (Figure \ref{fig:cmow}). Unlike the cubic binary carbides comprised of metal atoms from group 5 or 6, isostructural MoC and WC (Figure \ref{fig:cmow_graph}(a)) adopt the hexagonal $P\bar{6}m2$ (\#187) spacegroup, suggesting that Mo$_{1-x}$W$_x$C stoichiometry structures would be hexagonal as well. Examination of a $Imm2$ symmetry Mo$_{0.5}$W$_{0.5}$C phase that was 17~meV/atom above the convex hull (teal dot in Figure \ref{fig:cmow}) showed that it could not be derived from a decoration of a hexagonal metal-carbide lattice. Instead, it was related to a high-pressure phase of GaAs (\AFLOW\ prototype \href{http://aflow.org/prototype-encyclopedia/AB_oI4_44_a_b}{AB$\_$oI4$\_$44$\_$a$\_$b}) where the Ga atoms were replaced by C, half of the As atoms were substituted by Mo, and the other half by W. 

In addition to the elemental endpoints, as well as MoC and WC, cubic MoW and rhombohedral C$_2$Mo$_4$ (Figure \ref{fig:cmow}) comprise the \AFLOW\ convex hull for the CMoW system. Tetragonal Mo$_{14}$W$_2$ (labelled by a purple dot) lies 1~meV/atom above the hull -- a value that is within the error of the $k$-mesh and kinetic energy cutoffs employed in our planewave calculations. A different choice of \DFT\ functional, inclusion of zero point energy or finite temperature contributions may place this structure on the hull. We then examined if MTPs trained by \PRAPs\ on \AFLOW\ data could relax structures created with \textsc{RandSpg} and identify other thermodynamically stable compounds, not found within \AFLOW\ for the CMoW system.

We compare the \AFLOW\ hull with hulls predicted using the \PRAPs\ procedure (Figure \ref{fig:cmow} and Figures S9 and S10). At an \MTP\ level of 10 no additional structures emerged, and the \PRAPs\-derived hull was identical to the one found within \AFLOW. However, for an \MTP\ level of 16, additional thermodynamically stable structures were found, including $Imm2$ Mo$_{0.75}$W$_{0.25}$C, in addition to the on-and-near-hull compounds present in \AFLOW. The $Imm2$ phase, containing two formula units per primitive cell, resembles hexagonal MoC except that in every second layer half of the metal atoms are replaced by W, and the substituted metal-containing triangular nets are arranged in an $\cdot\cdot\cdot$ABAB$\cdot\cdot\cdot$ stacking sequence with respect to each other (Figure \ref{fig:cmow_graph}(b)). This phase likely originated from the \textsc{RandSpg} set, which was subsequently optimized, in an active learning sense, via the robust and accurate potentials. When an MTP of level 10 was used instead, $Imm2$ Mo$_{0.75}$W$_{0.25}$C was not on the convex hull likely because the relaxation process with the robust potential pushed this particular configuration too high in energy. To test this hypothesis the level 16 data for $Imm2$ Mo$_{0.75}$W$_{0.25}$C was concatenated with the structures that are present on the level 10 hull, and further analysis revealed that this phase was predicted to be thermodynamically stable.

In addition, four more structures, within 1~meV/atom of the convex hull, lay on the level 16 hull: $P\bar{6}m2$ Mo$_{0.5}$W$_{0.5}$C, $P\bar{6}m2$ Mo$_{0.333}$W$_{0.666}$C, $P\bar{6}m2$ Mo$_{0.666}$W$_{0.333}$C and $Cm$ Mo$_{0.25}$W$_{0.75}$C (Figure \ref{fig:cmow_graph}(c-f)). Though the first has the same composition as the structure present within \AFLOW, it is 17~meV/atom lower in energy. In fact, if we do not distinguish between the identities of the metal atoms, the \AFLOW\ structure can be transformed into $P\bar{6}m2$ Mo$_{0.5}$W$_{0.5}$C by doubling it along the $b$-axis followed by three sets of translations of various subsets of atoms. In both phases the C atoms fall within trigonal-prismatic holes, but in this particular structure the triangular (and square) faces all point along the same crystallographic direction, while in the \AFLOW\ structure half of the prisms are rotated, thereby swapping the axes along which the two sets of faces lie. Importantly, \PRAPs\-found $P\bar{6}m2$ Mo$_{0.5}$W$_{0.5}$C corresponds to a coloring of the hexagonal CMo/CW prototype structure with an $\cdot\cdot\cdot$ABAB$\cdot\cdot\cdot$ arrangement for the metal-containing hexagonal nets (Figure \ref{fig:cmow_graph}(c)). Similarly, the remaining three \PRAPs\-found structures can be derived from colorings of the hexagonal parent phase, with $P\bar{6}m2$ Mo$_{0.333}$W$_{0.666}$C and $P\bar{6}m2$ Mo$_{0.666}$W$_{0.333}$C being inverses of each other, while $Cm$ Mo$_{0.25}$W$_{0.75}$C can be described as a W-rich $\cdot\cdot\cdot$ABAB$\cdot\cdot\cdot$ layered decoration of this same hexagonal prototype. 

The identified near-and-on-hull phases lie on a straight line joining the two end-members comprising this CMo/CW series. They represent examples of an ensemble of phases with highly-variable concentrations, suggesting the existence of a solid solution with a very low critical temperature of the miscibility gap. To investigate this, we optimized $\sim$366 Mo$_{1-x}$W$_x$C structures ($x=$ $0.08\dot{3}$, $0.\dot{1}$, $0.125$, $0.1\dot{6}$, $0.\dot{2}$, $0.25$, $0.\dot{3}$, $0.375$, $0.41\dot{6}$, $0.\dot{4}$, $0.5$, $0.\dot{5}$, $0.58\dot{3}$, $0.625$, $0.\dot{6}$, $0.75$, $0.\dot{7}$, $0.8\dot{3}$) with 4-24 atoms in the unit cell, and between 2 and 86 unique structures were optimized per composition. The previously generated level 16 robust and accurate potentials were used to predict their energies and to relax them. Figure \ref{fig:cmow_graph}(g) plots the resulting enthalpies of formation, $\Delta H$, from the monocarbide endpoints: relaxed with the robust potential ({\small RR}), subsequently predicted by the accurate potential (\APRR), and finally relaxed with the accurate potential (\ARRR).  

All of the \DFT-optimized compounds fell on or within 5.7~meV/atom of the line joining the CMo and CW end points, suggesting that  their $\Delta H$ is close to 0~meV/atom.  For a given composition, various decorations were computed to be nearly isoenthalpic, suggesting that configurational entropy will play a role in the stability of this family of structures. Turning to the results obtained with the generated \MTPs, the computed $\Delta H$, as predicted on structures relaxed by the \RP, was largely positive (blue dots) with the deviation from the zero-energy line steadily increasing for larger W concentrations.  Whereas the distance from the CMo-CW tie-line, averaged over all structures, was calculated as being $0.83$~meV/atom ($\sigma=1.10$) via \DFT, the robust relaxed protocol resulted in an average tie-line distance of $78.39$~meV/atom ($\sigma=28.95$). Prediction of the energies of the robust relaxed structures with the \AP\ (green dots) yielded an average $\Delta H$ of $18.82$~meV/atom ($\sigma=14.96$). It is only via relaxation with the \AP\ (purple dots) that we obtain an average tie-line distance of $4.20$~meV/atom ($\sigma=3.58$). This example illustrates that structural relaxation with the \AP\ is key for obtaining energetics that are in good agreement with those derived from \DFT\ calculations.

\begin{figure*}
    \centering
    \includegraphics[width=0.95\textwidth]{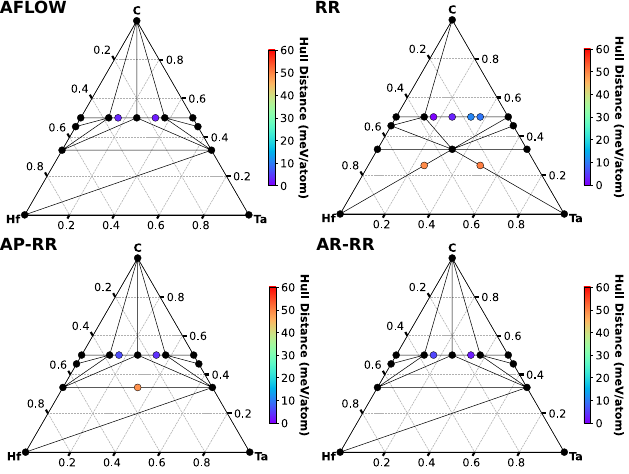}
    \caption{\small Structures within 60~meV/atom of the \AFLOW\ derived convex hull (top left), as well as the \PRAPs\ procedure at an \MTP\ level of 22 for the CHfTa system. Structures are colored (see color bar) according to the distances from the hull obtained after relaxing with the robust potential (RR), prediction of the enthalpies of the RR structures with the accurate potential (\APRR) and relaxing the RR structures with the accurate potential (\ARRR). Black dots are on the hull, and purple dots are within 1~meV/atom of the hull. In contrast to the plots shown in Figure \ref{fig:cmow}, the data illustrated here has not undergone further \DFT\ relaxations.}
    \label{fig:chfta}
\end{figure*}

The convex hulls discussed and presented above (Figure \ref{fig:cmow}) were optimized with \DFT, and the conclusions regarding thermodynamic stability of particular phases were made based upon these hulls. This procedure is common-place~\cite{GAML_Review} but it might make one wonder about the limits of the utility of \MLIAPs\ in \CSP. Part of the answer lies above where we show that ML can significantly reduce the number of required \DFT\ calculations. But, the other part of this answer is in the convex hull candidate structures: the output of \PRAPs\ relaxations and predictions before the final \DFT\ step. The analysis of the CMoW system suggested that relaxation with the \AP\ is key for obtaining energetics that are in-line with \DFT\ results. To further study this aspect, in Figure \ref{fig:chfta} we plot the convex hulls for the CHfTa system calculated at an \MTP\ level of 22. Comparison of the \AFLOW\ derived hull with one that is obtained after relaxation with the robust potential (RR) shows that the latter predicts a structure that is not found within \AFLOW, with  Hf$_{0.5}$TaC$_{0.5}$ composition, to lie on the hull (after \DFT\ relaxation, it falls 123~meV/atom above the hull) whereas Hf$_{1-x}$Ta$_x$C stoichiometries lie around 15~meV/atom above the hull. The rogue Hf$_{0.5}$TaC$_{0.5}$ structure disappears after \AP\ prediction, and the energies of the Hf$_{1-x}$Ta$_x$C species fall onto-and-just-above the hull. Relaxation with the \AP\ yields a hull that is virtually indistinguishable from the one derived from \AFLOW, similar to the results obtained for the CMoW system.  In the Supplementary Information we provide these same four convex hulls for each carbide system considered and each \MTP\ level, before and after subsequent relaxation with \DFT. Generally speaking, the structures that fall within 60~meV/atom of the convex hull, after being relaxed with the \RP\ followed by the \AP, also fall within 60~meV/atom of to finish These examples suggest that relaxations performed with an \AP\ will be useful as a further screening step that can be undertaken prior to performing expensive \DFT\ calculations in the materials prediction workflow.

\subsection{Discussion}

The density functional theory (\DFT) computed energies, forces and stresses found within the \AFLOW\ database of four ternary carbide systems (HfTaC, HfZrC, MoWC and TaTiC) were employed to train system specific machine learning interatomic potentials of the moment tensor potential (\MTP) flavor. A utility package that can be used to generate both robust potentials (\RP), capable of roughly relaxing any structure,  and accurate potentials (\AP), tailored towards the relaxation of low-energy structures, which was employed to automate this training, is described. The \AFLOW\ data was augmented with $\sim$6300 random symmetric structures resembling those that would be created in the first step of a crystal structure prediction (\CSP) search, and these were relaxed with \MTPs\ updated via active learning. For the HfTaC system, relaxation with the \AP\ yielded a convex hull that agreed perfectly with the one found within \AFLOW.  Moreover, this procedure identified five Mo$_{1-x}$W$_x$C stoichiometry compounds, not found within \AFLOW, that lay on the convex hull and corresponded to colorings of the hexagonal CMo/CW prototypes, illustrating how the described protocol can accelerate \CSP. Subsequently, the \RP\ and \AP\ were used to relax hundreds of Mo$_{1-x}$W$_x$C lattices spanning a broad composition range, and it was shown that relaxation with the \AP\ yielded formation enthalpies that were in excellent agreement with those computed via \DFT, and the resulting convex hull exhibits regions in which entropy plays a considerable role on the phase stability. The ideas and tools described here may aid in the generation of ML-IAPs from already existing \DFT\ data, to be used for materials prediction.

\section*{Methods}
\subsection{Computational Details}

The density functional theory (\DFT) calculations were performed using the Vienna \emph{ab initio} Simulation Package version 5.4.12~\cite{VASP} coupled with the Perdew, Burke, Ernzerhof ({\small PBE}) gradient-corrected exchange and correlation functional~\cite{PBE} and the projector augmented wave method~\cite{PAW}. During the active learning procedure, the {\small VASP} calculations were performed using $\Gamma$-centered Monkhorst-Pack $k$-meshes where the number of divisions along each reciprocal lattice vector was chosen such that the product of this number with the real lattice constant was 30~\AA{}. The carbon $2s^22p^2$, Hf $6s^25d^2$, Ta $6s^25d^3$, Zr $5s^24d^2$, Ti $4s^13d^3$, Mo $5s^24d^4$ and W $6s^25d^4$ electrons were treated as valence, and an energy cutoff of 400~eV was employed.  After training was complete, the convex hull analysis included a \DFT\ relaxation accomplished by calling \AFLOW's management protocol, using the standard settings described in Ref.\cite{AFLOWstd}; the \AFLOW\ hull data was also re-relaxed using this procedure. Structures from the \AFLOW\ database and those generated by \textsc{RandSpg}~\cite{RandSPG}, as described in the main text, comprised the full relaxation set employed for the development of the \MTPs. The crystals whose geometries were relaxed to construct Figure \ref{fig:cmow_graph} (g) were generated from $P\bar{6}m2$ Mo$_{0.5}$W$_{0.5}$C  using the Supercell program, employing the merge option to remove duplicate structures ~\cite{Supercell}. 

\PRAPs\ was run on each ternary carbide using \MTP\ levels 10, 16, and 22 with a \MLIP-relaxation-iteration limit of 100, and an extrapolation grade of $2< \gamma \le 10$. The cutoff distances for generating the \MTP\ were 1.1~\AA{} $< x <$ 5~\AA{}.  Active learning was, in most cases, declared to be converged when no additional structures were considered for addition to the training set. In the case of the level 22 trainings, the ALS procedure was stopped when the number of structures to be added to the training set was less than 1\% of the number already in the training set. \PRAPs\ filtered out configurations with interatomic distances 1.1~\AA{} $< d <$ 3.1~\AA{} before beginning the pre-training, and when beginning the \AP\ training removed all structures that were higher than 50~meV/atom of the most stable configuration for each composition. \\

\noindent{\bf Code Availability.}
The \PRAPs\ code will be released in a subsequent publication, and in the meanwhile, may be obtained from the corresponding authors upon reasonable request. \\[2ex]

\noindent\textbf{Data Availability} 
The datasets generated during and/or analyzed during the current study are summarized in the supplementary information, and are available from the corresponding author on reasonable request. \\[2ex]

\noindent\textbf{Acknowledgements}
We would like to gratefully acknowledge the DoD {\small SPICES MURI} sponsored by the Office of Naval Research (Naval Research contract N00014-21-1-2515) for their financial support of this work. Calculations were performed at the Center for Computational Research at SUNY Buffalo (http://hdl.handle.net/10477/79221). We would like to acknowledge Xiaoyu Wang for artistic help and guidance, Masashi Kimura for assistance with the convex hull diagrams, Hagen Eckert, Xiomara Campilongo and Corey Oses for fruitful discussions. \\[2ex]

\noindent\textbf{Author Contributions}
E.Z.\ and S.C.\ conceived the research and supervised the study. J.R.\ carried out the method development of the \PRAPs\ code, and performed the calculations and analysis. All authors participated in discussing the results, and commented on the manuscript. \\[2ex]

\noindent\textbf{Additional Information} \\
\textbf{Supplementary information} accompanies the paper on the \emph{npj Computational Materials} website.

\noindent\textbf{Competing Interests:} The authors declare no competing interests.

%\newpage
\bibliographystyle{naturemag} %EOPAPER
%PhysRevwithTitles_DOI_v1b
\normalbaselines %Fixes spacing of bibliography %EOPAPER

%\bibliographystyle{apsrev4-1}
%\bibliography{PRAPs_refs_ordered,duke}

\end{document}

\date{}
\maketitle

\tableofcontents

\clearpage
\newpage

\section{Plan for Robust and Accurate Potentials (PRAPs) Training}

\begin{table*}[ht]
\centering
\begin{tabular}{ |c|c|c|c|c| }
 \hline
 Level 10 & Pre-RP & RP & AP-RR & AP-Low \\ 
 \hline \hline
 CHfTa & 6-0/7-0 & 6-0/9-0 & 6-0/8-0 & 6-0/10-0 \\
 \hline
 CHfZr & 3-3/7-0 & 2-0/9-0 & 2-0/9-0 & 2-0/9-0 \\
 \hline
 CMoW & 8-0/8-0 & 6-0/9-0 & 3-0/10-0 & 6-0/3-1 \\
 \hline
 CTaTi & 7-0/7-0 & 4-0/4-0 & 4-0/8-0 & 4-0/7-0 \\
 \hline \hline 
 Level 16 & Pre-RP & RP & AP-RR & AP-Low \\
 \hline \hline 
 CHfTa & 6-0/8-0 & 6-0/10-0 & 6-0/8-0 & 6-0/10-0 \\
 \hline
 CHfZr & 3-2/7-0 & 6-0/10-0 & 2-0/9-0 & 6-0/9-0 \\
 \hline
 CMoW & 8-0/4-0 & 7-0/10-0 & 7-0/8-0 & 7-0/1-1 \\
 \hline
 CTaTi & 8-0/6-0 & 7-0/4-0 & 6-0/9-0 & 7-0/8-0 \\
 \hline \hline
 Level 22 & Pre-RP & RP & AP-RR & AP-Low \\
 \hline \hline
 CHfTa & 3-2/6-0 & 3-0/9-0 & 6-0/10-0 & 3-0/10-0 \\
 \hline
 CHfZr & 1-2/6-0 & 2-0/9-0 & 2-0/7-0 & 2-0/9-0 \\
 \hline
 %CMoW & 8-0/8-0 & 2-0/9-0 & 2-0/3-1 & 2-0/1-0 \\
 %\hline
 CTaTi & 3-3/6-0 & 4-0/7-0 & 5-0/9-0 & 4-0/9-0 \\
 \hline
\end{tabular}
\caption{The number of low energy configurations correctly and incorrectly predicted (A-B) and the number of high energy configurations correctly and incorrectly predicted (C-D) by PRAPs (denoted as A-B/C-D) for the ternary carbide systems studied at different MTP levels of theory. The Pre-Robust Potential (Pre-RP) and the Robust Potential (RP) are compared against the original AFLOW data (after removing structures with unphysically short interatomic distances). The Accurate Potential (AP) is used to make two predictions and two comparisons. First it is compared against the low-energy robust-relaxed data (AP-RR), and then compared against the low-energy AFLOW data (AP-Low).} 
\label{table:highlow}
\end{table*}

\section{Training Data and Root Mean Square Errors}

\begin{figure*}
    \centering
    \includegraphics[width=\textwidth]{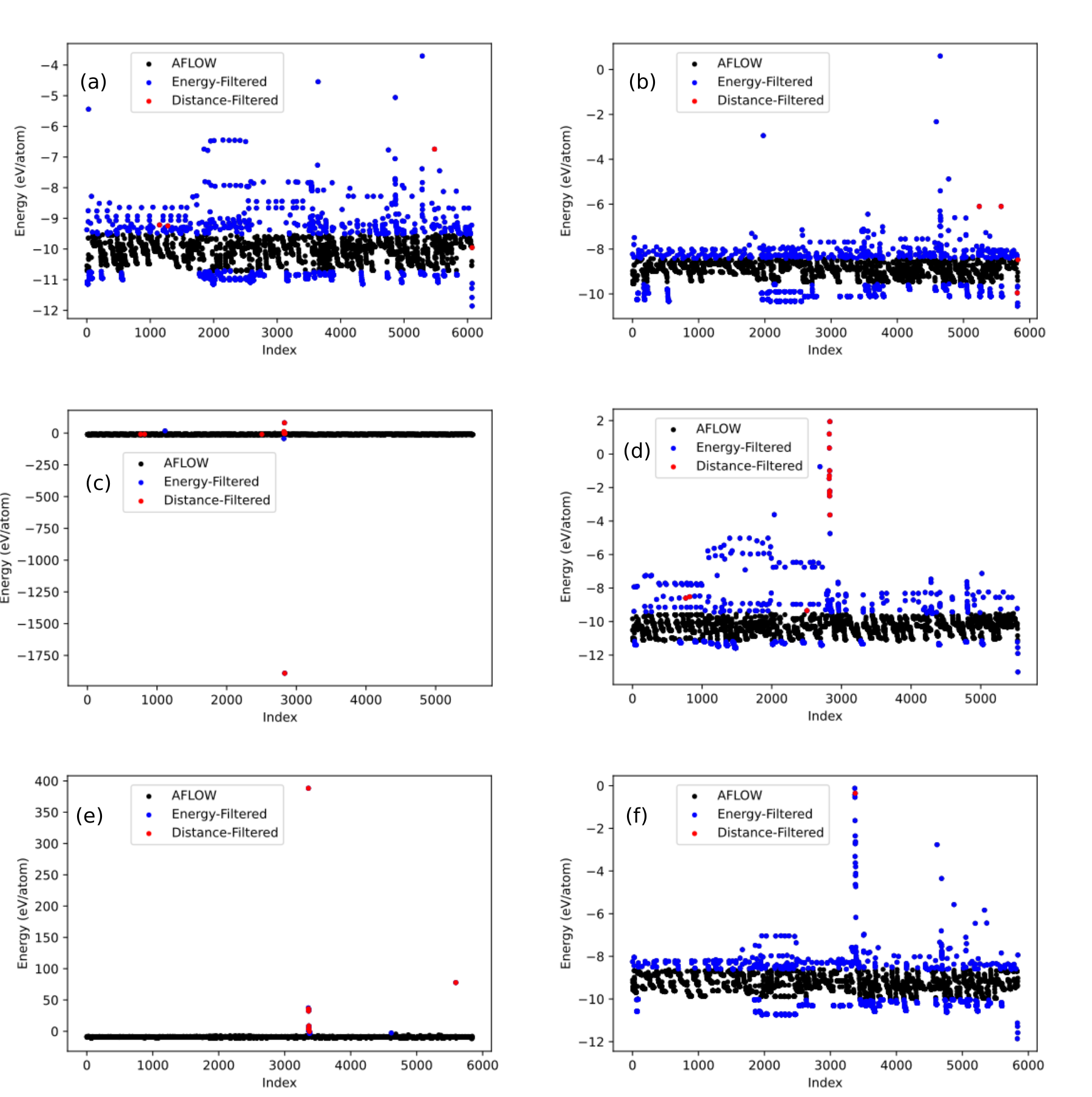}
    \caption{Energy values found in the original AFLOW data for: (a) CHfTa, (b) CHfZr, (c and d) CMoW raw and zoomed-in, and (e and f) CTaTi raw and zoomed-in. Configurations not removed by filtering are marked in black. Configurations removed by the distance filtering are marked in red. Configurations removed during testing of energy filtering are marked in blue.}
    \label{fig:aflow-e}
\end{figure*}

To better understand the factors contributing to the root mean square errors (RMSE), we plotted the DFT energies of the structures present in the AFLOW training set versus an arbitrarily chosen structure index in Figure \ref{fig:aflow-e}. Configurations with interatomic distances outside of the range $1.1$ \AA $< x < 3.1$ \AA, were automatically filtered from the training set. While a majority of the configurations present possessed energies that fell within $-9.5 \pm 1.0$~eV/atom, $5.6\%$ fell outside this range, some by up to -40~meV/atom, and $30\%$ fell outside of one standard deviation from the mean. \\

\noindent
PRAPs always filters structures by distance, but we considered the possibility of adding energy as a secondary filtering condition. After filtering by distance the remaining filtrate was filtered by only including configurations with energies within one standard deviation from the mean. A new robust potential (Level 10e) was generated from this filtered data and plots were generated to investigate the effect of the filtering on the energy predictions, appearing below in Figure \ref{fig:lev10e}.

\begin{figure*}
    \centering
    \includegraphics[width=\textwidth]{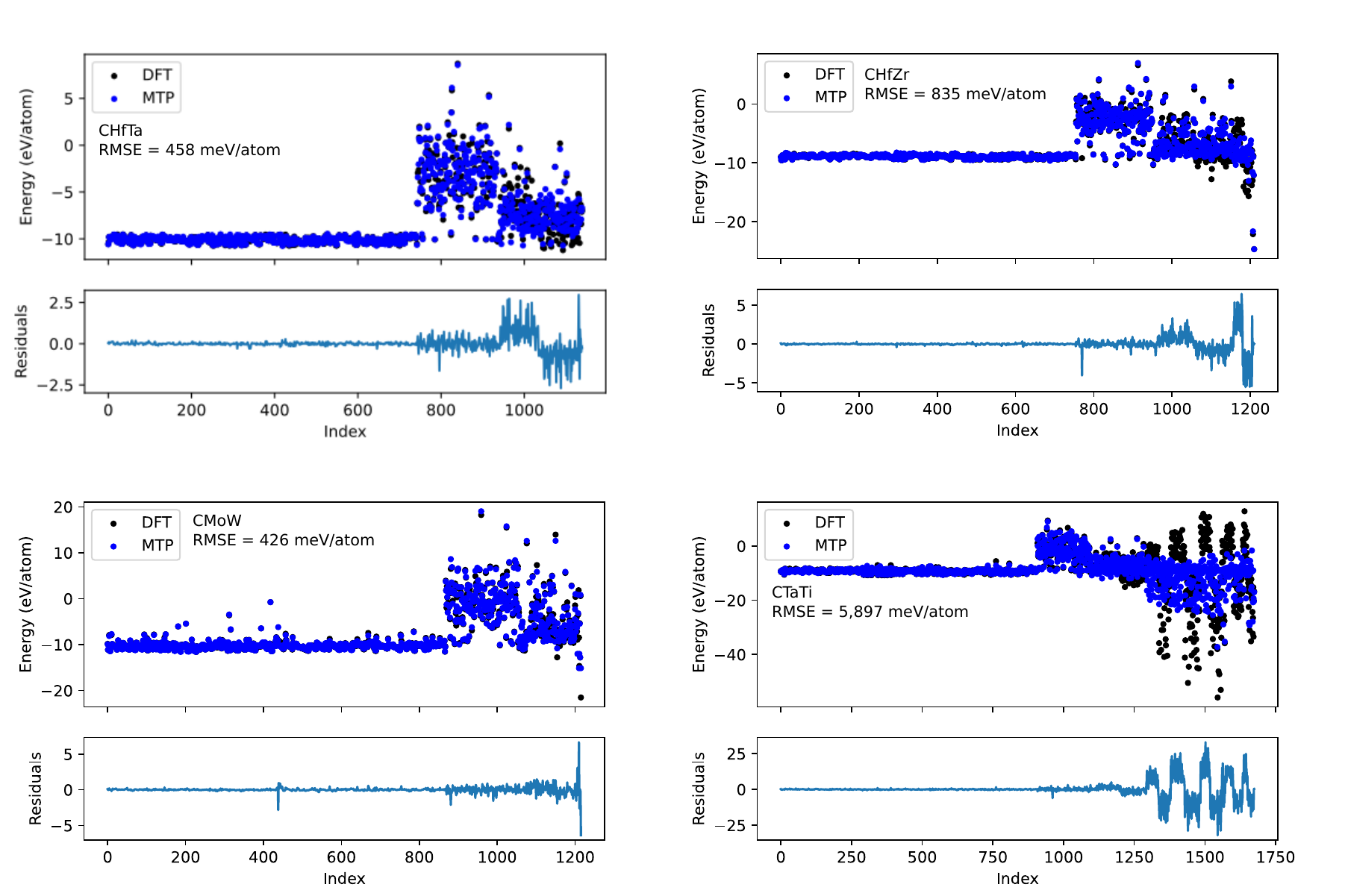}
    \caption{The AFLOW data (Figure \ref{fig:aflow-e}, all dots) was filtered by both distance and energy (Figure \ref{fig:aflow-e}, black dots) and submitted for training. Here we show the energies of the Robust Potential training set, as calculated by DFT and the Robust Potential (MTP, Level 10e). The training set contains two regions: the relatively flat region in the first half of the x-axis contains configurations straight from the AFLOW data found in Figure \ref{fig:aflow-e}, the other region in the second half of the x-axis with the widely spread data contains configurations generated during the active learning procedure, which may include configurations that are not chemically sensible or converged, reflecting the quality of the generating MTP. The associated root-mean-square-errors appear in each plot (compare to 197, 302, 448, and 704 meV/atom, respectively, from the main text, Table II) and the differences between the DFT and Robust Potential appear below as residual traces.}
    \label{fig:lev10e}
\end{figure*}

\newpage
\noindent
To gain a better understanding of the predictive power of our MTPs and the origin of the previously-stated RMSEs, we examined the differences between the original AFLOW data and the trained potential's prediction. In Table \ref{table:percents}, we counted how many configurations had differences within one standard deviation of the mean and how many configurations had differences less-than 0.5 eV/atom. This was done for each system, for both Robust and Accurate Potentials, using the standard Level 10 data (see main text) and the energy-filtered data from above (Level 10e).

\begin{table*}[ht]
\centering
\begin{tabular}{ |c|c|c|c|c| }
 \hline
 Level 10 & \multicolumn{2}{|c|}{RP} & \multicolumn{2}{|c|}{AP} \\
 \hline
  & Stdev & $<0.5$ & Stdev & $<0.5$ \\
 \hline
 CHfTa & 77.9\% & 84.3\% & 87.4\% & 99.6\% \\
 \hline
 CHfZr & 81.6\% & 80.5\% & 90.9\% & 99.3\% \\
 \hline
 CMoW & 81.3\% & 79.1\% & 86.4\% & 98.9\% \\
 \hline
 CTaTi & 85.7\% & 81.1\% & 87.3\% & 99.7\% \\
 \hline \hline
 Level 10e & \multicolumn{2}{|c|}{RP} & \multicolumn{2}{|c|}{AP} \\
 \hline
  & Stdev & $<0.5$ & Stdev & $<0.5$ \\
 \hline
 CHfTa & 77.2\% & 75.3\% & 87.0\% & 98.7\% \\
 \hline
 CHfZr & 79.2\% & 73.4\% & 89.9\% & 99.9\% \\
 \hline
 CMoW & 79.7\% & 78.5\% & 91.2\% & 97.7\% \\
 \hline
 CTaTi & 78.3\% & 55.4\% & 92.2\% & 72.4\% \\
 \hline
\end{tabular}
\caption{Percent of the training set configurations whose predicted energies differed by less-than one standard deviation of the mean and less-than 0.5 eV/atom from the original AFLOW data.}
\label{table:percents}
\end{table*}

\noindent
We performed a set of tests on each of the four systems at level 10 to test the effects of structural diversity. In these, we replaced the set of ~6000 RandSPG structures of diverse space groups, with a set of ~200 HCP, FCC, and BCC structures from RandSPG. The energy RMS values are reported in meV/atom in Table \ref{table:nicelats} with the original values from Table II of the main text in parentheses for comparison. 

\begin{table*}[ht]
\centering
\begin{tabular}{ |c|c|c|c|c|c| }
 \hline
 System & \multicolumn{3}{|c|}{Training Errors (meV/atom)} & \multicolumn{2}{|c|}{Prediction Errors (meV/atom)} \\
 \hline
  & Pre-trained & RP & AP & RP & AP \\
 \hline
 CHfTa & 38 (34) & 186 (197) & 162 (179) & 77 (77) & 53 (70) \\
 \hline
 CHfZr & 32 (31) & 244 (302) & 95 (136) & 76 (92) & 42 (89) \\
 \hline
 CMoW & 66 (69) & 458 (448) & 152 (187) & 489 (490) & 80 (142) \\
 \hline
 CTaTi & 43 (41) & 803 (704) & 279 (167) & 251 (237) & 150 (90) \\
 \hline
\end{tabular}
\caption{Root mean square errors of the energy for training and prediction across the four systems where the lattices of the RandSPG structures were restricted to HCP, FCC, and BCC only. Original values from Table II of the main text are in parentheses.}
\label{table:nicelats}
\end{table*}

\newpage
\noindent
Finally, we performed a small test at level 10 across all four systems to examine the usefulness of the pre-training procedure. The entire PRAPs training procedure was conducted, omitting the pre-training step, and the timings and errors were compared with those obtained from the original tests as described in the main text. The final DFT calculations and convex hull analysis were omitted as that time can be separated from the training time. The results are provided in Table \ref{table:alstests}. Omitting the pre-training gave significant improvements in speed at the cost of MTP quality, as measured by energy RMS error (values in parentheses are from the original run, Table II in the main text). Wall times (D-HH:MM:SS) were provided instead of CPU times for a more faithful estimate of run time. 

\begin{table*}[ht]
\centering
\begin{tabular}{ |c|c|c|c|c|c|c| }
 \hline
 System & \multicolumn{2}{|c|}{Wall Time} & \multicolumn{2}{|c|}{Training Errors (meV/atom)} & \multicolumn{2}{|c|}{Prediction Errors (meV/atom)} \\
 \hline
  & Original & No Pre-train & RP & AP & RP & AP \\
 \hline
 CHfTa & 7-20:50:57 & 6-00:08:55 & 622 (197) & 109 (179) & 183 (77) & 105 (70) \\
 \hline
 CHfZr & 8-10:27:47 & 8-02:14:44 & 613 (302) & 94 (136) & 160 (92) & 136 (89) \\
 \hline
 CMoW & 7-12:24:07 & 4-21:53:13 & 760 (448) & 201 (187) & 504 (490) & 434 (142) \\
 \hline
 CTaTi & 12-05:49:09 & 6-02:22:11 & 1104 (704) & 296 (167) & 329 (237) & 188 (90) \\
 \hline
\end{tabular}
\caption{Comparison of PRAPs performance with and without the pre-training step. Times are given as Days-Hours:Minutes:Seconds. Error values are reported in meV/atom. The parentheses contain the errors for the control group at Level 10 with pre-training as reported in the main text, Table II.}
\label{table:alstests}
\end{table*}

\newpage
\section{Convex Hulls}

\begin{figure*}
    \centering
    \includegraphics[width=\textwidth]{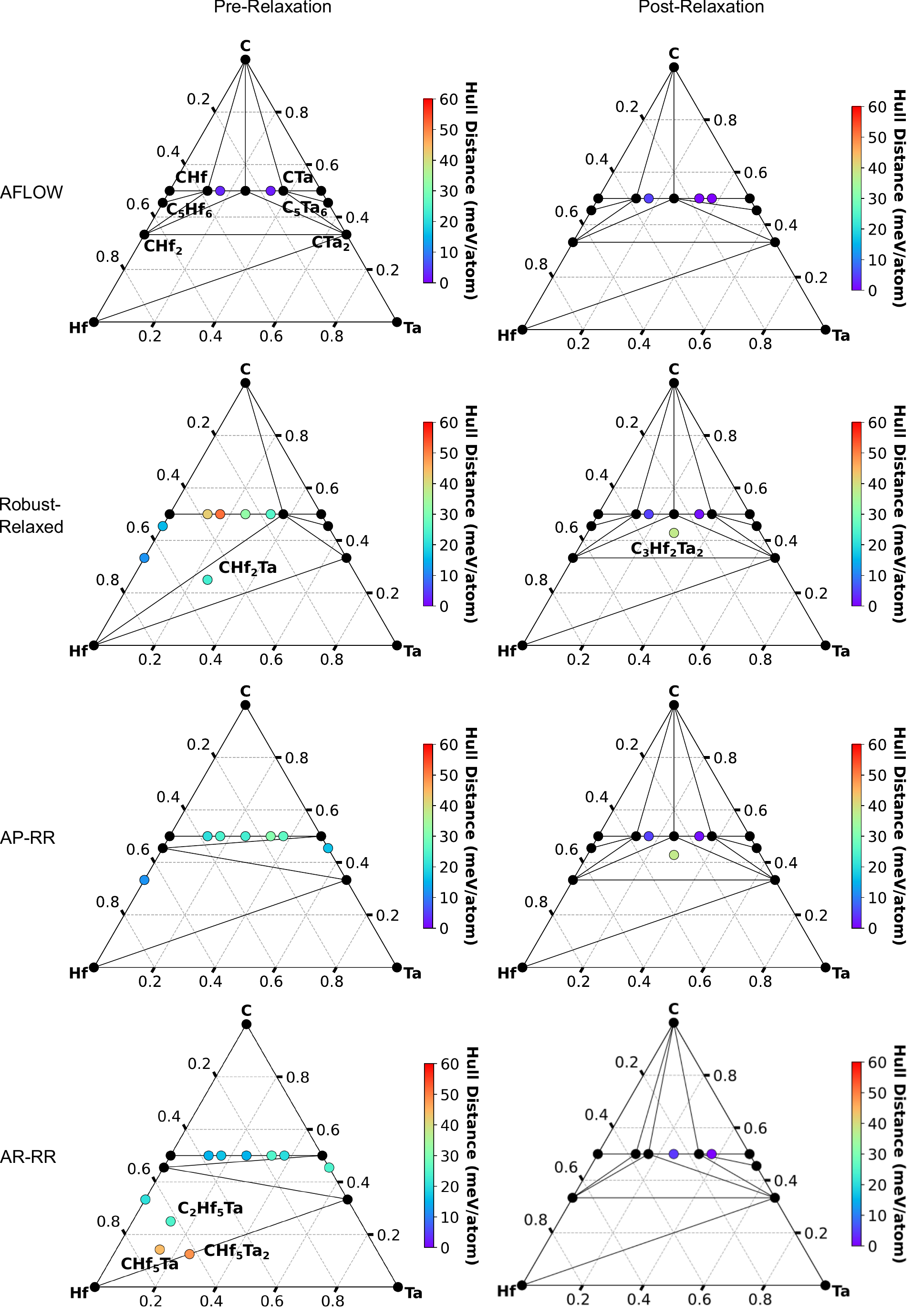}
    \caption{Convex Hulls of CHfTa at Level 10. The unlabeled structures on the line between CHf and CTa are, from left-to-right, C$_4$Hf$_3$Ta, C$_3$Hf$_2$Ta, C$_2$HfTa, C$_3$HfTa$_2$, and C$_4$HfTa$_3$.}
    \label{fig:chfta10}
\end{figure*}

\begin{figure*}
    \centering
    \includegraphics[width=\textwidth]{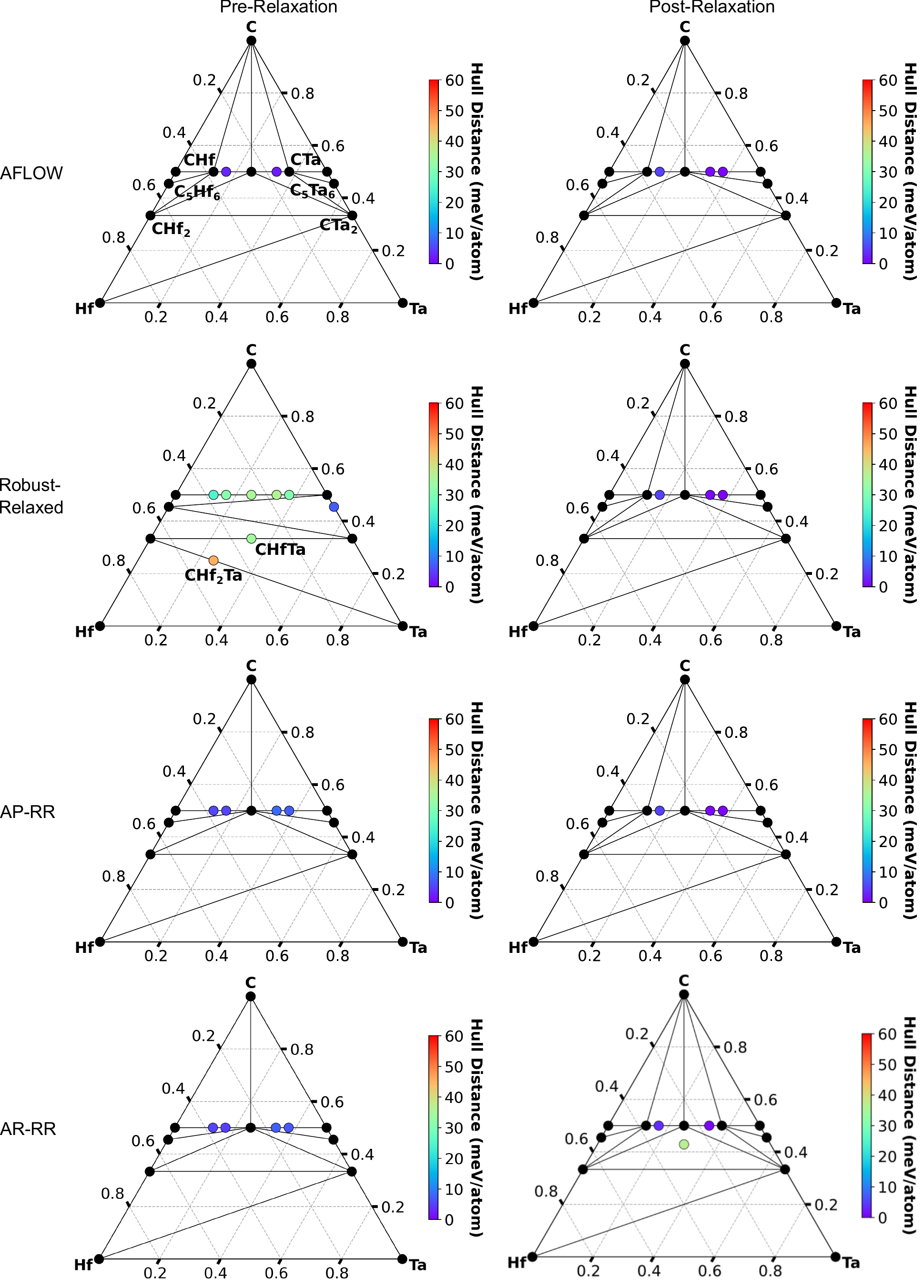}
    \caption{Convex Hulls of CHfTa at Level 16. The unlabeled structures on the line between CHf and CTa are, from left-to-right, C$_4$Hf$_3$Ta, C$_3$Hf$_2$Ta, C$_2$HfTa, C$_3$HfTa$_2$, and C$_4$HfTa$_3$.}
    \label{fig:chfta16}
\end{figure*}

\begin{figure*}
    \centering
    \includegraphics[width=\textwidth]{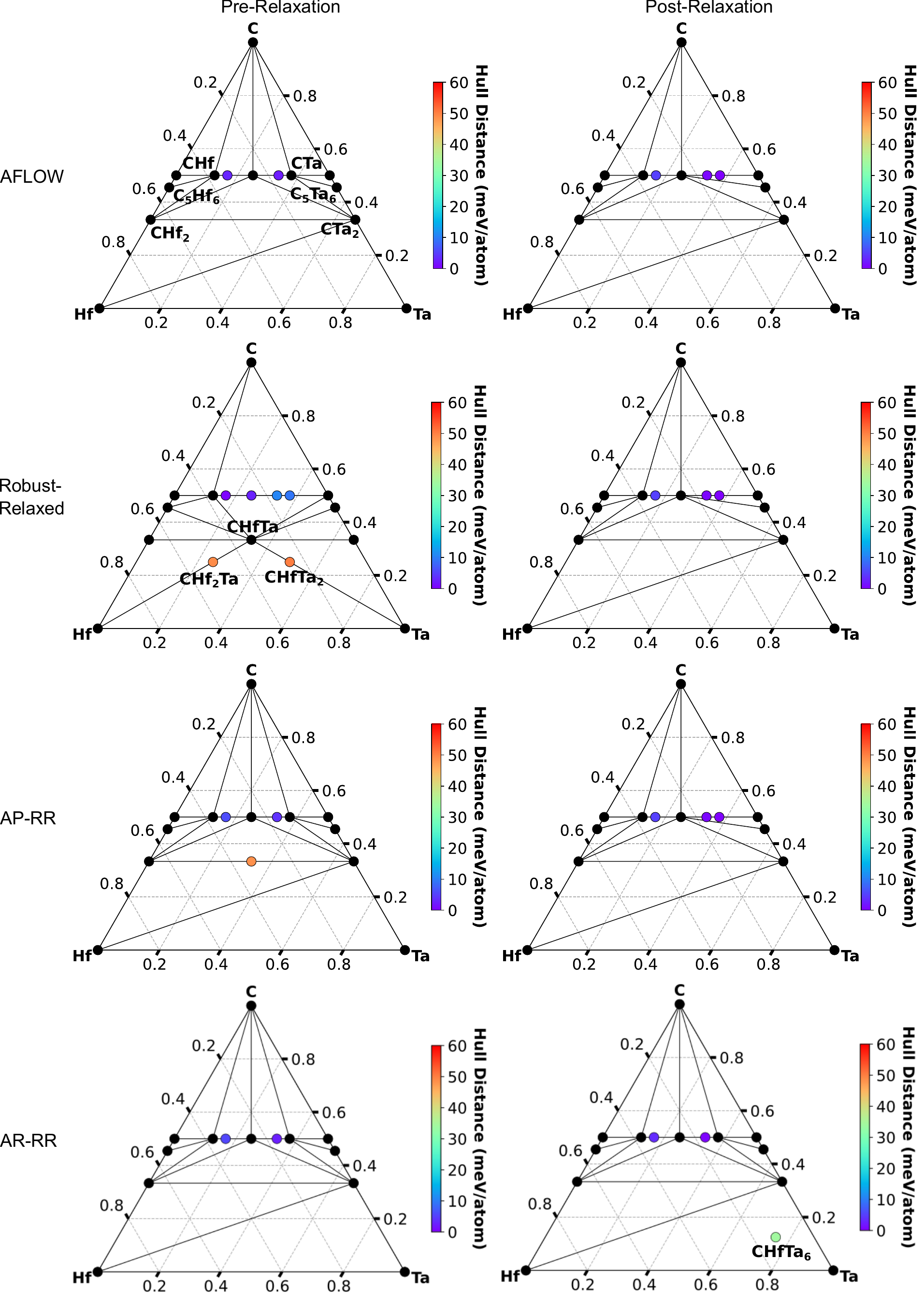}
    \caption{Convex Hulls of CHfTa at Level 22. The unlabeled structures on the line between CHf and CTa are, from left-to-right, C$_4$Hf$_3$Ta, C$_3$Hf$_2$Ta, C$_2$HfTa, C$_3$HfTa$_2$, and C$_4$HfTa$_3$.}
    \label{fig:chfta22}
\end{figure*}

\begin{figure*}
    \centering
    \includegraphics[width=\textwidth]{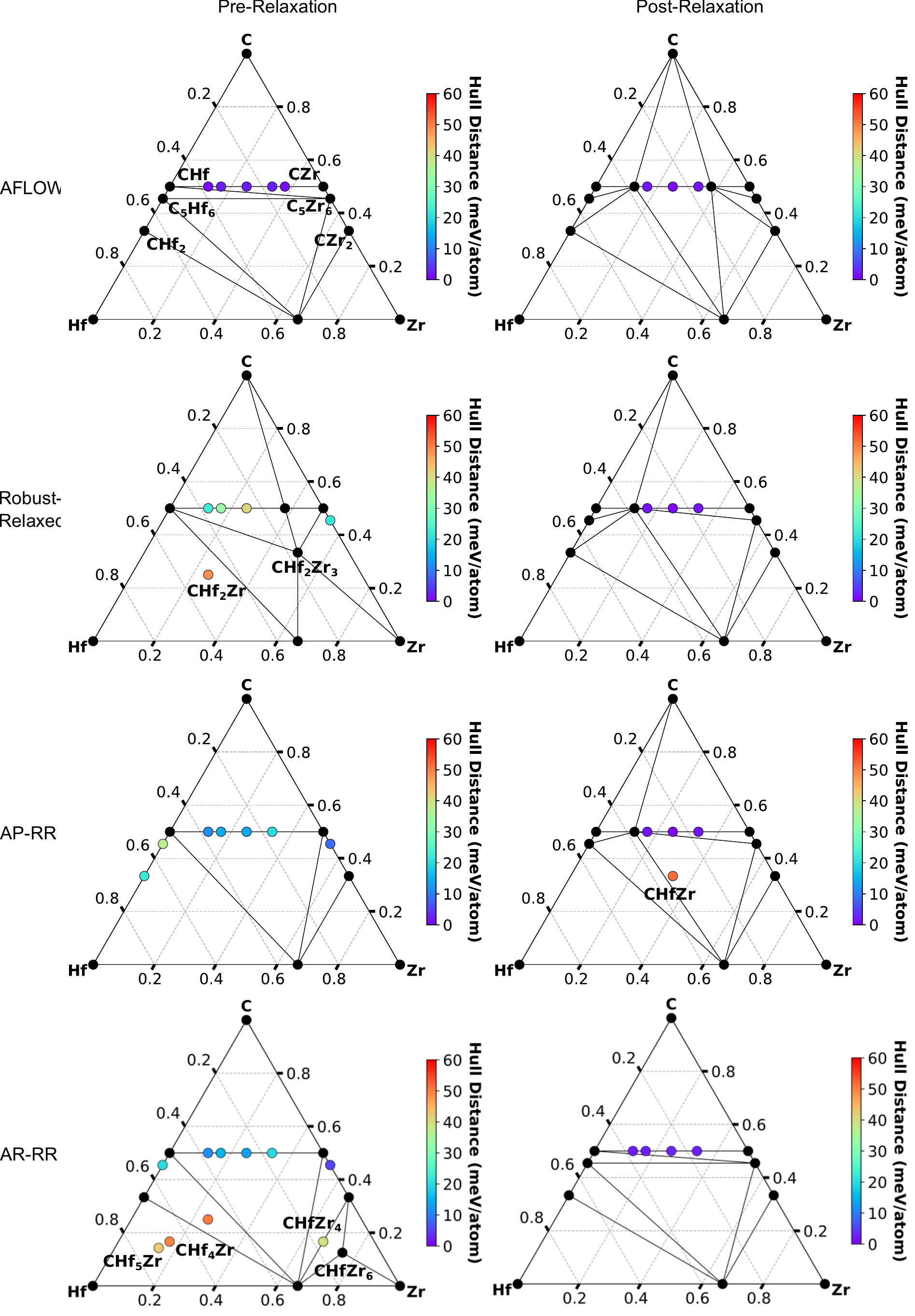}
    \caption{Convex Hulls of CHfZr at Level 10. The unlabeled structures on the line between CHf and CZr are, from left-to-right, C$_4$Hf$_3$Zr, C$_3$Hf$_2$Zr, C$_2$HfZr, C$_3$HfZr$_2$, and C$_4$HfZra$_3$.}
    \label{fig:chfzr10}
\end{figure*}

\begin{figure*}
    \centering
    \includegraphics[width=\textwidth]{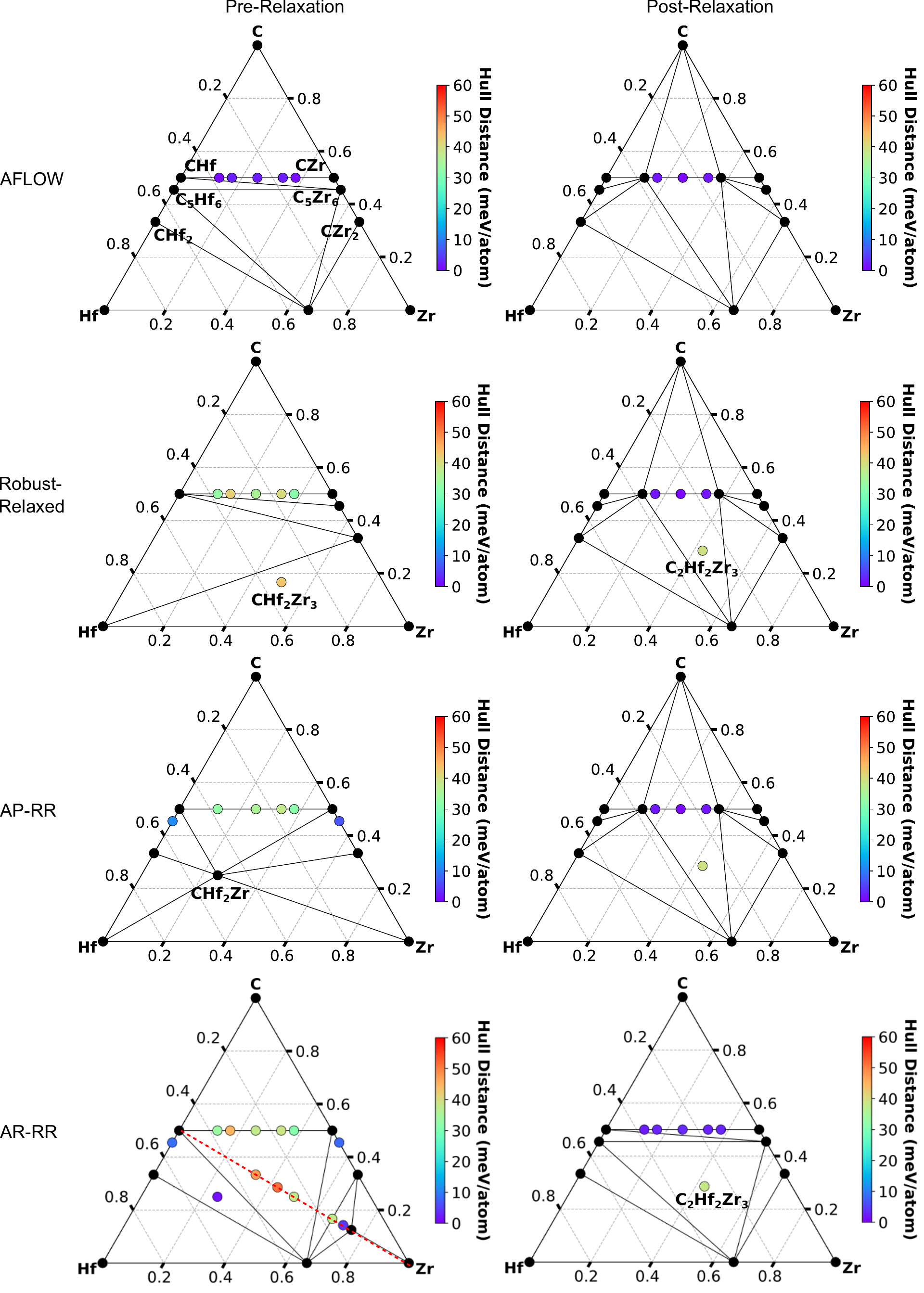}
    \caption{Convex Hulls of CHfZr at Level 16. The unlabeled structures on the line between CHf and CZr are, from left-to-right, C$_4$Hf$_3$Zr, C$_3$Hf$_2$Zr, C$_2$HfZr, C$_3$HfZr$_2$, and C$_4$HfZr$_3$.}
    \label{fig:chfzr16}
\end{figure*}

\begin{figure*}
    \centering
    \includegraphics[width=\textwidth]{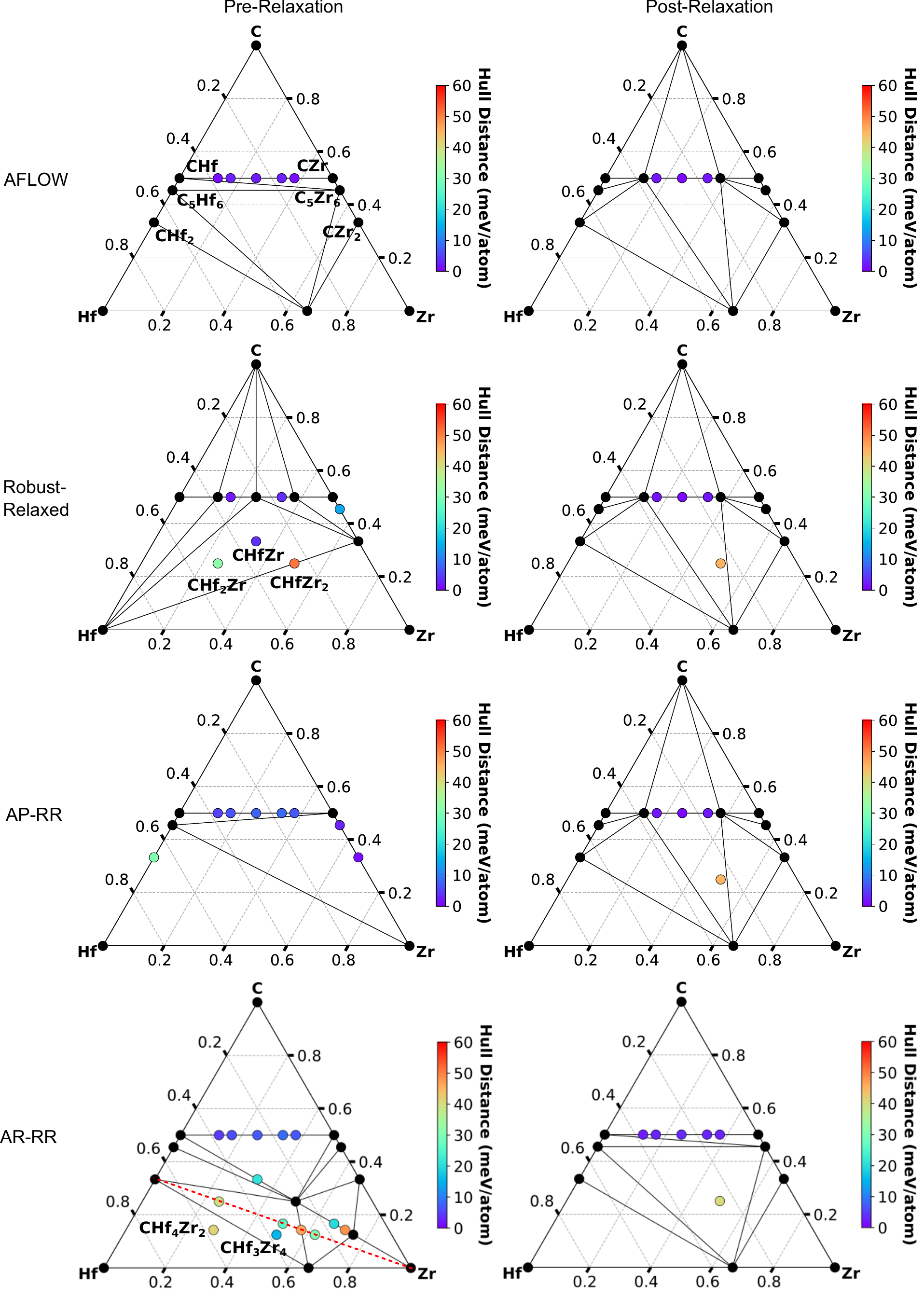}
    \caption{Convex Hulls of CHfZr at Level 22. The unlabeled structures on the line between CHf and CZr are, from left-to-right, C$_4$Hf$_3$Zr, C$_3$Hf$_2$Zr, C$_2$HfZr, C$_3$HfZr$_2$, and C$_4$HfZr$_3$.}
    \label{fig:chfzr22}
\end{figure*}

\begin{figure*}
    \centering
    \includegraphics[width=\textwidth]{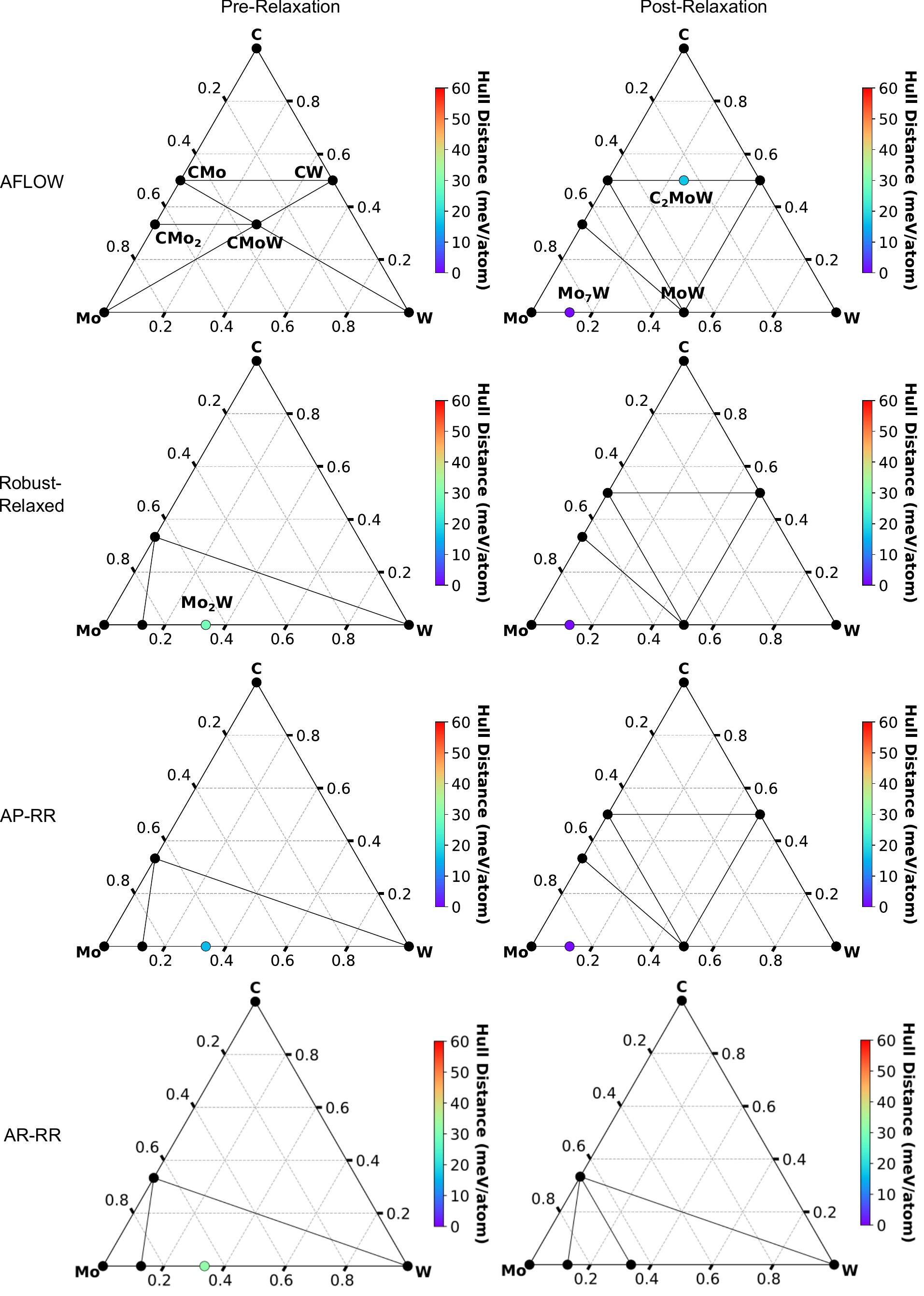}
    \caption{Convex Hulls of CMoW at Level 10.}
    \label{fig:cmow10}
\end{figure*}

\begin{figure*}
    \centering
    \includegraphics[width=\textwidth]{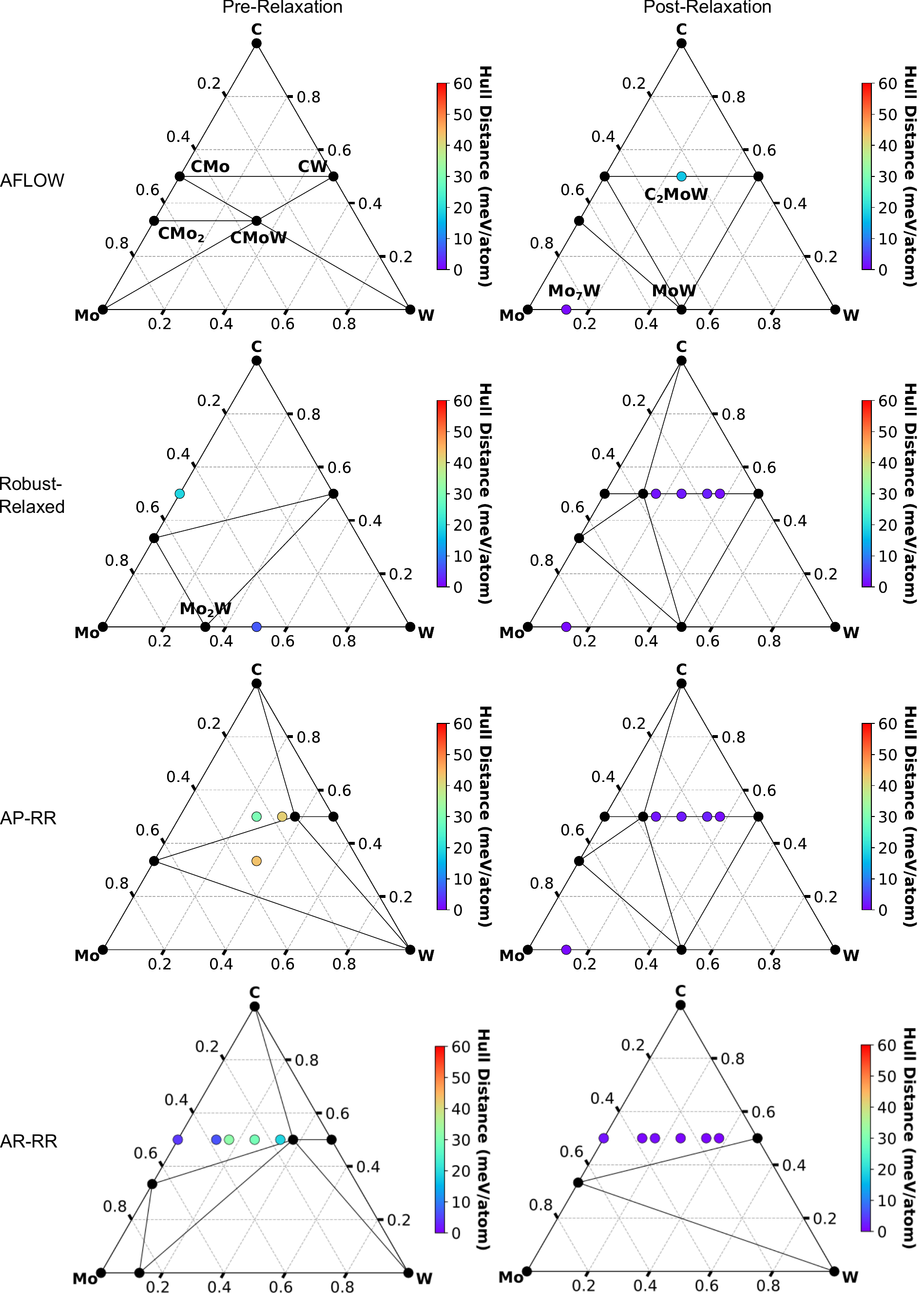}
    \caption{Convex Hulls of CMoW at Level 16. The unlabeled structures on the line between CMo and CW are, from left-to-right, C$_4$Mo$_3$W, C$_3$Mo$_2$W, C$_2$MoW, C$_3$MoW$_2$, and C$_4$MoW$_3$.}
    \label{fig:cmow16}
\end{figure*}

\begin{figure*}
    \centering
    \includegraphics[width=\textwidth]{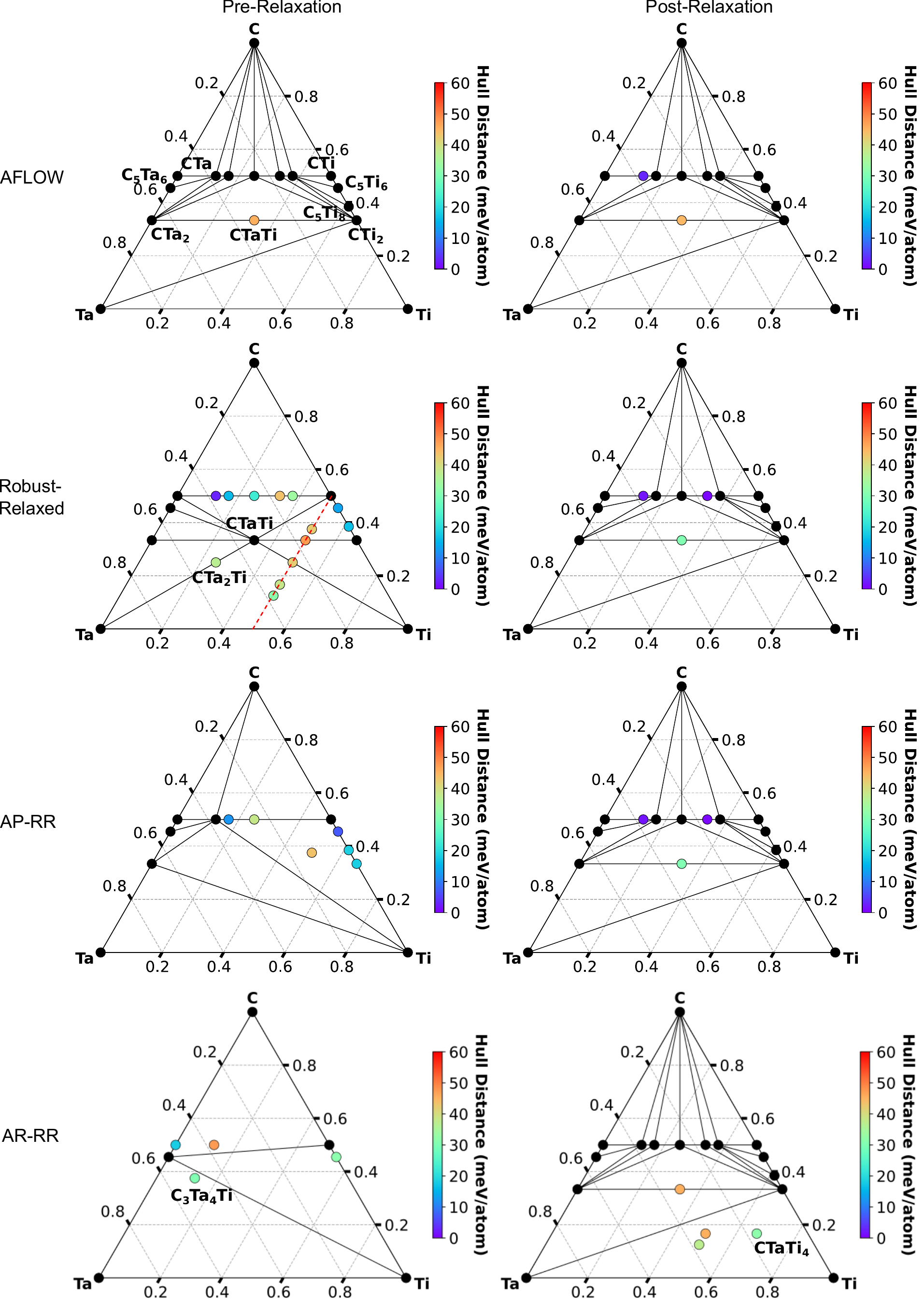}
    \caption{Convex Hulls of CTaTi at Level 10. The unlabeled structures on the line between CTa and CTi are, from left-to-right, C$_4$Ta$_3$Ti, C$_3$Ta$_2$Ti, C$_2$TaTi, C$_3$TaTi$_2$, and C$_4$TaTi$_3$. The unlabeled structures on the dotted red line between CTi and TaTi are, from top-to-bottom, C$_3$TaTi$_4$, C$_2$TaTi$_3$, CTaTi$_2$, CTa$_2$Ti$_3$, and CTa$_3$Ti$_4$.}
    \label{fig:ctati10}
\end{figure*}

\begin{figure*}
    \centering
    \includegraphics[width=\textwidth]{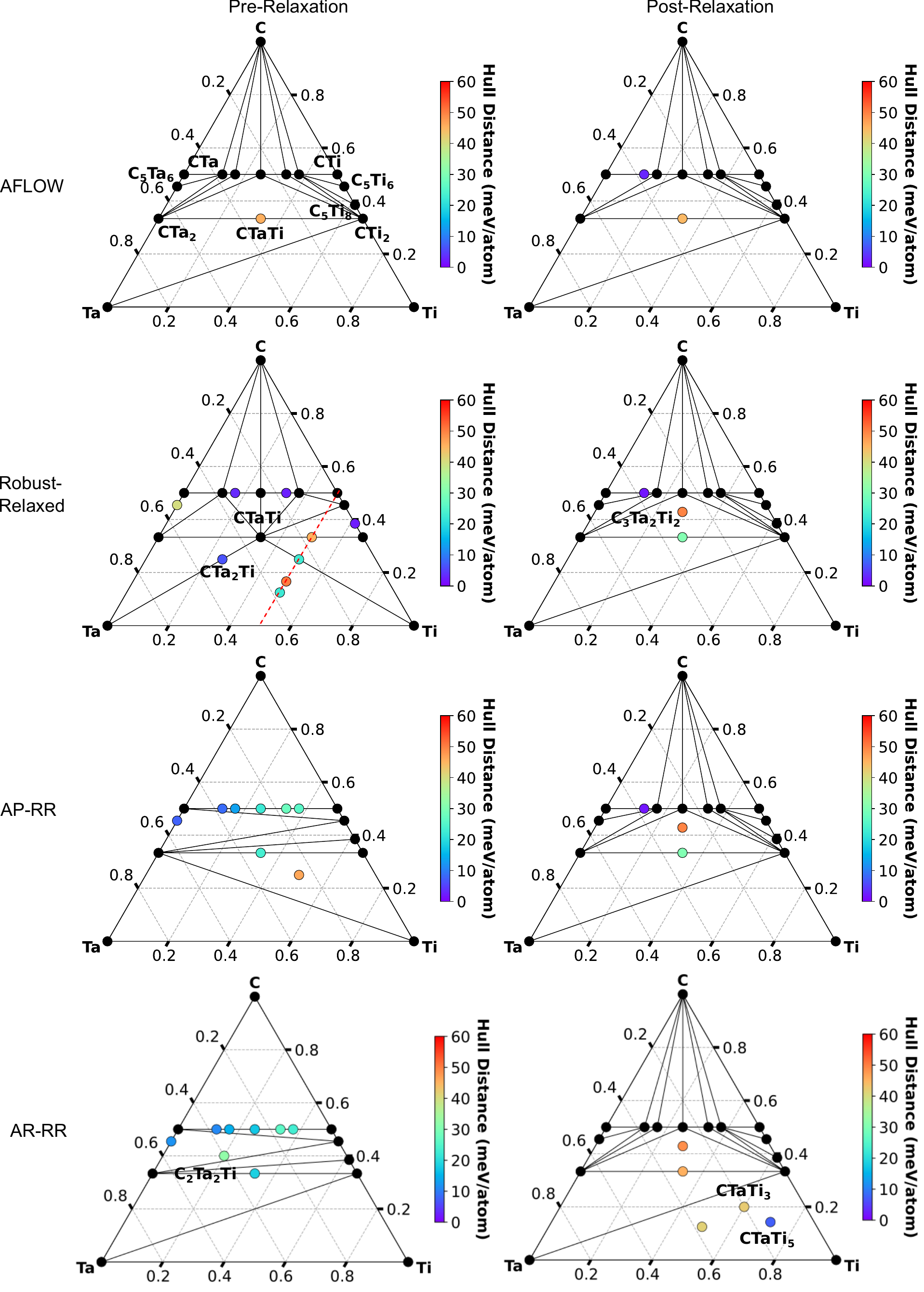}
    \caption{Convex Hulls of CTaTi at Level 16. The unlabeled structures on the line between CTa and CTi are, from left-to-right, C$_4$Ta$_3$Ti, C$_3$Ta$_2$Ti, C$_2$TaTi, C$_3$TaTi$_2$, and C$_4$TaTi$_3$. The unlabeled structures on the dotted red line between CTi and TaTi are, from top-to-bottom, C$_3$TaTi$_4$, C$_2$TaTi$_3$, CTaTi$_2$, CTa$_2$Ti$_3$, and CTa$_3$Ti$_4$.}
    \label{fig:ctati16}
\end{figure*}

\begin{figure*}
    \centering
    \includegraphics[width=\textwidth]{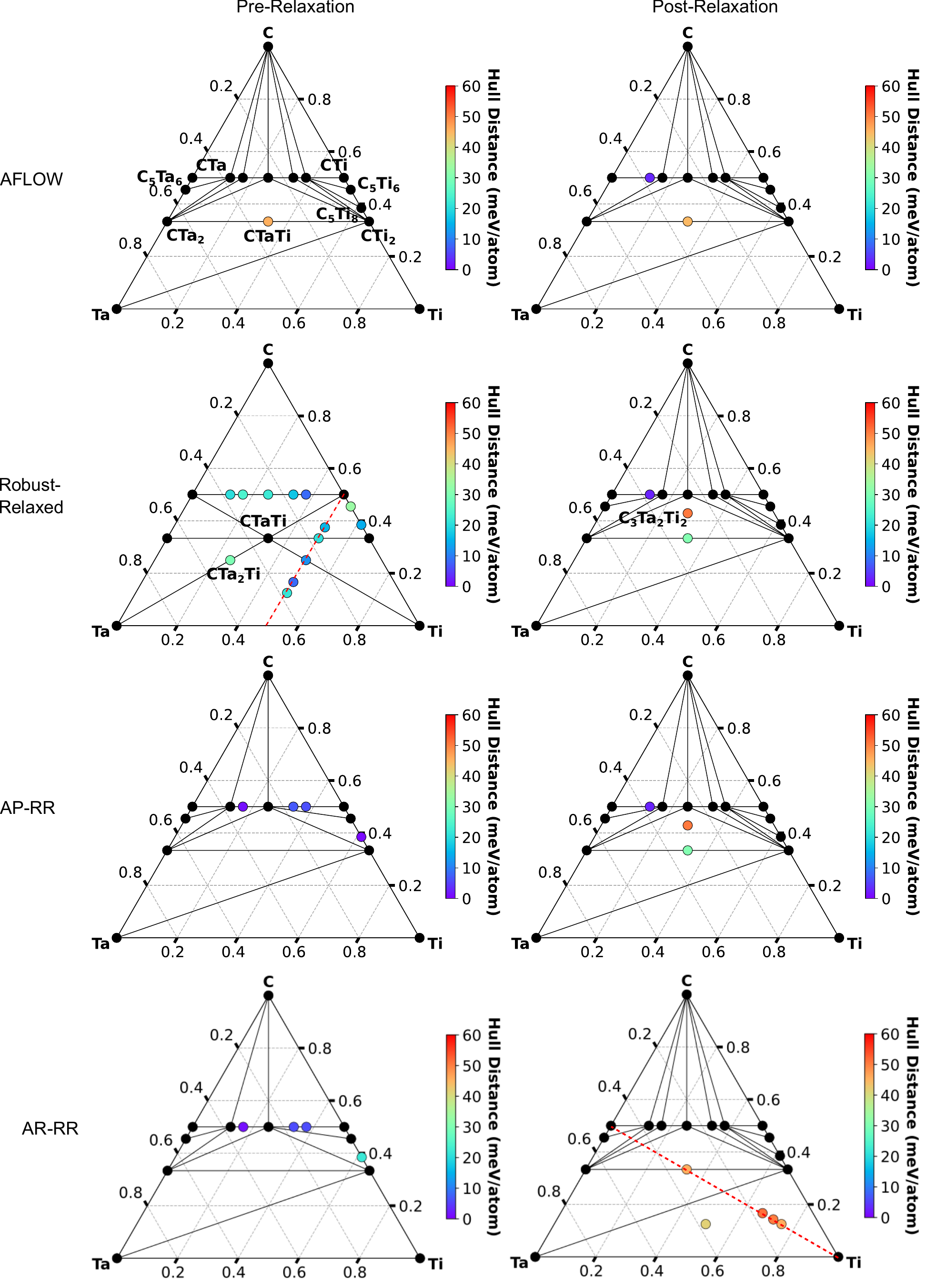}
    \caption{Convex Hulls of CTaTi at Level 22. The unlabeled structures on the line between CTa and CTi are, from left-to-right, C$_4$Ta$_3$Ti, C$_3$Ta$_2$Ti, C$_2$TaTi, C$_3$TaTi$_2$, and C$_4$TaTi$_3$. The unlabeled structures on the dotted red line between CTi and TaTi are, from top-to-bottom, C$_3$TaTi$_4$, C$_2$TaTi$_3$, CTaTi$_2$, CTa$_2$Ti$_3$, and CTa$_3$Ti$_4$.}
    \label{fig:ctati22}
\end{figure*}

This section contains a subset of the convex hulls calculated by PRAPs. If desired by the user, PRAPs will calculate eleven convex hulls: two for the DFT data before-and-after relaxation, and nine using the Robust and Accurate Potentials. Six of the hulls are presented here in a table-like format, organized into rows and columns from Figures \ref{fig:chfta10} to \ref{fig:ctati22}. The AFLOW data (top row) is combined with the \textsc{RandSPG} data and filtered to remove unphysical structures as described in the main text. This filtered set is then relaxed using the Robust Potential to form the Robust-Relaxed set (RR, second row). Then the RR set is filtered to obtain only configurations within 50 meV/atom of the lowest energy configuration for each composition and the Accurate Potential is used to relax these structures (AP-RR, bottom row). The Accurate Potential is also used on the AFLOW data to determine if it can reproduce the AFLOW convex hull (AP-v, not shown). The left column contains hulls generated from the MLIP output while the right shows the hulls of the same data after DFT relaxation via AFLOW's standard procedure. To calculate the remaining hulls, PRAPs will concatenate two files together: the Post-Relaxation AFLOW convex hull (shown below in the top-right of each figure) and the Post-Relaxation Accurate Potential hulls (shown below in the middle-right and lower-right of each figure). This allows a direct comparison of the ``literature" and ``experiment" and will tell the user whether or not structures predicted by PRAPs are still on the convex hull when considering the original AFLOW data. These hulls are not shown below as almost all of the hulls remain unchanged from those already shown.

\newpage

